\documentclass[12pt, letterpaper, twoside]{article}

\usepackage{authblk}
\usepackage{graphicx}
\usepackage{dcolumn}
\usepackage{bm}
\usepackage{amsmath}
\usepackage{amssymb}

\usepackage{siunitx}
\usepackage{tabularx} 
\newcolumntype{b}{X}
\newcolumntype{s}{>{\hsize=.27\hsize}X}
\usepackage[version=4]{mhchem}

\usepackage{ulem}

\usepackage{xcolor}
\definecolor{mypink}{RGB}{255, 0, 181}
\definecolor{myblue}{RGB}{0, 149, 255}
\definecolor{mycyan}{RGB}{0, 255, 249}
\definecolor{mydarkgreen}{RGB}{0, 128, 128}
\definecolor{mygreen}{RGB}{0, 128, 0}

\begin{document}
\title{Cooperativity-dependent folding of single-stranded DNA}

\author[1,2,*]{X. Viader-Godoy}
\author[3]{C. R. Pulido}
\author[3]{B. Ibarra}
\author[1]{M. Manosas}
\author[1]{F.Ritort}
\affil[1]{%
 Departament de F\'{i}sica de la Mat\`eria Condensada, Universitat de Barcelona, Diagonal 647, 08028 Barcelona, Spain
}%
\affil[2]{%
 Department of Physics and Astronomy, University of Padova, via Marzolo 8, 35131 Padova, Italy
}%

\affil[3]{
Instituto Madrileño de Estudios Avanzados en Nanociencia, IMDEA Nanociencia. 28049 Madrid, Spain
}%

\affil[*]{\textit{ xavier.viadergodoy@unipd.it}}

\date{\today}

\maketitle             
             
\begin{abstract}
The folding of biological macromolecules is a fundamental process of which we lack a full comprehension. Mostly studied in proteins and RNA, single-stranded DNA (ssDNA) also folds, at physiological salt conditions, by forming non-specific secondary structures that are difficult to characterize with biophysical techniques. Here we present a helix-coil model for secondary structure formation, where ssDNA bases are organized in two different types of domains (compact and free). The model contains two parameters: the energy gain per base in a compact domain, $\epsilon$, and the cooperativity related to the interfacial energy between different domains, $\gamma$. We tested the ability of the model to quantify the formation of secondary structure in ssDNA molecules mechanically stretched with optical tweezers. The model reproduces the experimental force-extension curves in ssDNA of different molecular lengths and varying sodium and magnesium concentrations. Salt-correction effects for the energy of compact domains and the interfacial energy are found to be compatible with those of DNA hybridization. The model also predicts the folding free energy and the average size of domains at zero force, finding good agreement with secondary structure predictions by Mfold. We envision the model could be further extended to investigate native folding in RNA and proteins. 
\end{abstract}

\section{Introduction}

Single-stranded nucleic acids participate in a myriad of processes \cite{lodishmolecular2008,albertsmolecular2018}. Bases in single-stranded DNA (ssDNA) and ssRNA tend to form base pairs of different geometries and stabilities, from the most stable Watson-Crick to the weaker Hoogsteen or wobble base pairs \cite{Nagaswamy}. Base pairs are stabilized by hydrogen bonds and stacking interactions \cite{Florian1999,SPONER200143} that induce high-order tertiary structures \cite{RNApseudoknots05,Varshney2020}. Biologically active ssRNAs fold into compact domains, stabilized by base-pairing interactions, connected by flexible unpaired regions facilitating RNA conformational changes \cite{Wan2011}. Although DNA is found in a double-stranded form in the cell genome, ssDNA is generated as an essential intermediary during every aspect of DNA metabolism, replication \cite{prokaryoticReplication1992,reviewReplication2002,replication2010}, transcription  \cite{eukariotictranscription1991}, and repair \cite{dnarepair2005, SMEtranscription2005, STRACY2014}. The tendency of ssDNA to form secondary structures interferes with and modulates the action of DNA processing enzymes.

Understanding how molecular interactions drive molecular folding is a fundamental problem in biology, from RNA \cite{Thirumalai2005} to proteins \cite{Dill2012} but also in naturally occurring \cite{Portella2010,Woodside2015} and synthetic \cite{Schneider2019} DNA structures. After decades of research, conflicting views on how nucleic acids and proteins fold remain \cite{Woodside2015,Bryngelson1995,Wallace2003,Mahen2010,Karplus2011,Englander2014}. Besides the key biological role of ssDNA, understanding how ssDNA folds might shed light on the general mechanisms behind molecular folding in nucleic acids and proteins due to the ubiquitous presence of hydrogen-bonded and stacking interactions.

Secondary structure formation in nucleic acids can be investigated with pulling experiments with optical and magnetic tweezers \cite{Neuman2008}, which allow stretching individual molecules by applying mechanical forces in the pN range and measuring their force-extension curves (FECs)  \cite{SMITH1996,Croquette2001,BUHAL2004}. 
These experiments show that ssDNA behaves as a semiflexible polymer, with FECs described using ideal elastic polymer models (such as the Worm-Like Chain -WLC- model \cite{Bustamante1994}). However, at low forces ($\lesssim15$ pN) FECs deviate from the ideal elastic behavior, showing consistent compaction in the form of a force shoulder \cite{hairpinssecstruc2001,bosco2014elastic}. This compaction indicates that ssDNA folds via secondary structure formation quantified by the reduction in the molecular extension.

We introduce a helix-coil model for ssDNA folding. In the model, bases along the ssDNA chain belong to two types of domains, compact (C) and free (F). C-domains contain bases forming secondary structure, whereas F-domains contain unfolded free bases. Originally proposed by Zimm-Bragg and Lifson for protein folding \cite{zimm1959theory,lifson1961theory}, the helix-coil model has been extended to DNA hybridization \cite{poland1970theory,fixman1977theory,kafri2000dna,garel2004generalized,badasyan2005helix} and RNA folding \cite{pagnani2000glassy}. Surprisingly though, and to the best of our knowledge, the helix-coil model has never been used before for non-specific secondary structure formation in DNA and RNA.

We performed pulling experiments with optical tweezers on individual ssDNA molecules over a wide range of experimental conditions, varying molecular length and ionic strength, to test our model. The model reproduces the FECs of individual ssDNA molecules over two-three decades of molecular lengths and sodium and magnesium salt concentrations. In comparison with more complex models \cite{hairpinssecstruc2001, Muller2002,einert2011secondary}, the one presented here is described in terms of two parameters: the energy per base forming C-domains ($\epsilon$) and the interfacial energy between adjacent domains (cooperativity $\gamma$). 
Previous studies addressed the effect of salt on secondary structure formation in ssDNA \cite{Croquette2001,bosco2014elastic}, yet the effect of molecular length has never been addressed. The helix-coil model permits us to determine the finite-size corrections to the free energy of formation of folded ssDNA and the average size of domains at zero force. Despite its simplicity, the model is compatible with Mfold secondary structure predictions, showing its power for predicting cooperativity-dependent folding of ssDNA. 
\section{\label{S:Materials} Materials and Methods}
\subsection{\label{subsec:OT} Optical tweezers setup}
Experiments were carried out using a miniaturized dual-beam optical tweezers setup described in \cite{Smith2003,gieseler2020optical}. Briefly, two tightly focused counter-propagating laser beams (P=200 mW, $\lambda$=845 nm) created  a single optical trap. A DNA molecule was tethered between two micron-sized polystyrene beads, one captured in the optical trap and the other held on top of a glass micropipette [Fig.~\ref{fig:1}(a)].
dsDNA handles were labeled with Biotin or Digoxigenin (See Sec. S1 in the Supplemental Material at [URL will be inserted by publisher] for details) to bind selectively to Streptavidin (2.1 $\mu$m Kisker Biotech) and anti-Digoxigenin coated beads (3.0-3.4 $\mu$m Kisker Biotech). The experiments were performed in a microfluidics chamber. The force exerted on the optically trapped bead was determined by directly measuring the change of light momentum using Position Sensitive Detectors (PSDs). The position of the optical trap was determined by diverting $\sim$8\% of each laser beam to a secondary PSD. The instrument has a resolution of 0.1 pN and 1 nm at a 1 kHz acquisition rate.

\subsection{DNA substrates}
ssDNA easily adsorbs non-specifically on surfaces which makes it difficult to manipulate and to control the number of ssDNA nucleotides that are effectively stretched. In order to avoid this problem, we used DNA hairpins that were unfolded mechanically to generate ssDNA (see Sec.~\ref{subsec:molecular_extension} for details). In addition, the position at which mechanical unwinding of the hairpin starts can be used as a fiducial marker for alignment of the force-extension curves (FECs, see below). We used eight hairpins, with different sequences and lengths spanning from $\sim 120$b to $\sim 14$kb, named $\textrm{H}_N$, $N$ being the number of ssDNA bases released by the hairpin unfolding. We performed two different preparations for the DNA substrates (See Fig. S1 in the Supplemental Material at [URL will be inserted by publisher] for details).
The shorter hairpins ($120$b and $204$b) were synthesized by annealing and ligating a different set of oligonucleotides (See Sec. S2 in the Supplemental Material at [URL will be inserted by publisher] for their sequences). The longer hairpins (from $700$b to $13680$b) were synthesized using a long DNA fragment obtained either by a PCR amplification or by a digestion of the linearized $\lambda$-phage DNA. All hairpins were annealed to 29bp dsDNA handles that were used as spacers. Labeling of the handles was achieved by a Digoxigenin/Biotin tailing using a terminal transferase (See Sec. S1 in the Supplemental Material at [URL will be inserted by publisher] for details).

\subsection{\label{subsec:buffers} Buffers}

 Experiments with ssDNA of different lengths (Figs.~\ref{fig:1} and \ref{fig:3}) were performed at  10mM Tris-HCl pH7.5, 0.01\%\ce{NaN3} and 10mM \ce{MgCl2}. For varying \ce{NaCl} and \ce{MgCl2} concentration, previous data from our lab \cite{bosco2014elastic} was used: i) sodium buffer (10mM Tris-HCl pH7.5, EDTA 1mM, 0.01\%NaN$_3$) and NaCl  concentrations (10, 25, 50, 100, 250, 500 and 1000mM); and ii) magnesium buffer (10mM Tris-HCl pH7.5, 0.01\%\ce{NaN3}) and  \ce{MgCl2} concentrations (0.5, 1, 2, 4 and 10mM). To determine the ideal elastic response of ssDNA we also performed tests in glyoxal at 1M concentration (See App.~\ref{S:Ap.glyoxal}). All data was obtained at \SI{25}{\celsius}.

\subsection{\label{subsec:molecular_extension}ssDNA generation and determination of its molecular extension}

\begin{figure}[!ht]
\centering\includegraphics[width=\textwidth]{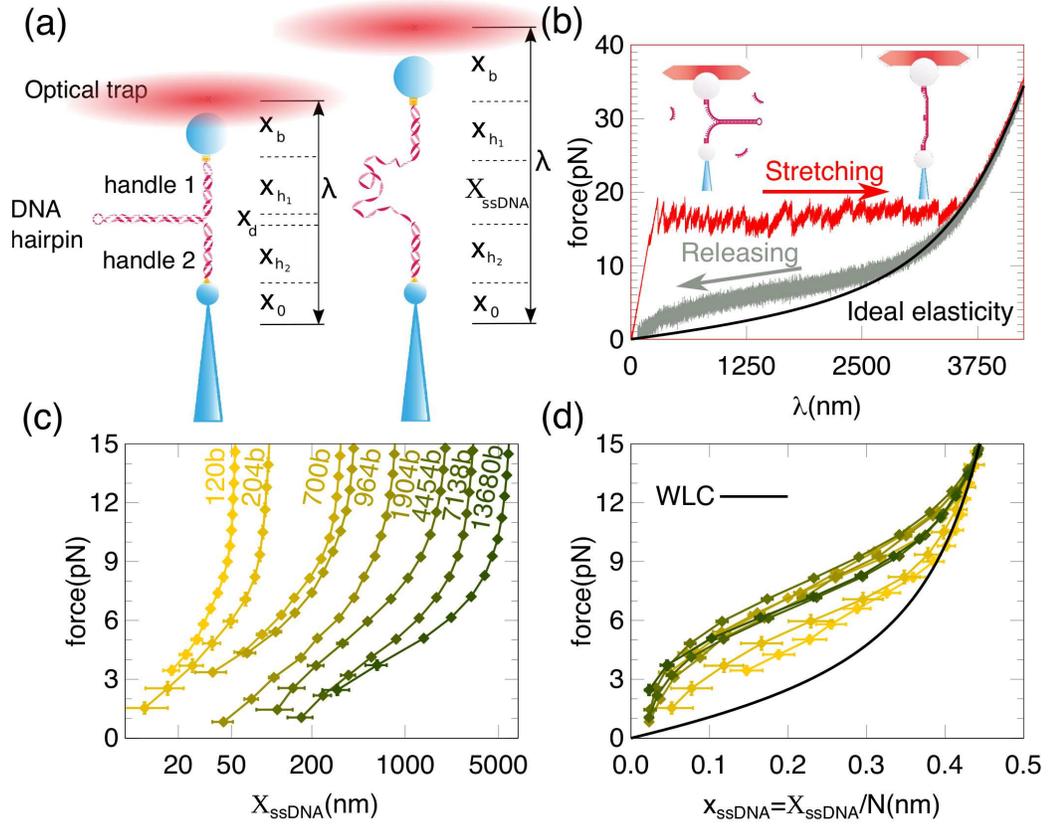}
\caption{\label{fig:1} \scriptsize{{\bf Experimental setup and ssDNA force-extension curves.}
(a) Sketch of the experimental setup for generating ssDNA. A DNA hairpin, bearing a functionalized dsDNA handles at each strand is tethered between two micron-sized beads; by moving the optical trap the molecule is stretched and the unfolding of the hairpin is observed at $f\sim15$ pN (b) Stretching (red) and releasing (gray) force-distance curves of the hairpin H$_{7138}$ in the presence of an oligonucleotide complementary to the loop region. Binding of the oligonucleotide blocks the rezipping of the hairpin, leading to the generation of ssDNA (Sec. \ref{subsec:molecular_extension}) and secondary structure formation (grey curve). The black line shows the elastic WLC model (Sec. \ref{subsec:elastic_models}). (c) ssDNA FECs obtained for the different hairpins (lengths ranging from 120b to $\sim$14000b, from yellow to dark green) by converting trap-distance ($\lambda$) into molecular extension ($X_{\textrm{ssDNA}}$) (Sec.~\ref{subsec:molecular_extension}). The x axis is represented in log scale. (d) FECs in (c) normalized by the number of bases ($N$). Color code as in panel (c). The secondary structure emerges at forces below $\sim10$ pN, as a shoulder deviating from the WLC [Eq.(\ref{eq:wlc_marko_siggia})] (black line). For $N \gtrsim 500$ FECs collapse into a single curve. The force shoulder for the 204b and 120b molecules is systematically lower, showing finite-length effects. Error bars were obtained by averaging 5-10 releasing cycles of 5-10 different tethers for each hairpin. Experiments were performed at \SI{25}{\celsius} in 10 mM \ce{MgCl2}.}}
\end{figure}

Several methods have been developed for obtaining ssDNA \cite{EMBO2009,Huguet2010PNAS, candellitoolbox2013, mcintosh2014}. In this work, ssDNA substrate was generated from a DNA hairpin using the blocking oligonucleotide method described in Ref.~\cite{bosco2014elastic}. Initially, the DNA hairpin was tethered between the two polystyrene beads. An oligonucleotide of $\sim 25-30$b, complementary to the loop region of the hairpin, was flowed into the chamber (See Table S2 in the Supplemental Material at [URL will be inserted by publisher] for the sequences). Above 15 pN the hairpin was fully unzipped [Fig.~\ref{fig:1}(b)], allowing the annealing of the complementary oligonucleotide to the exposed hairpin-loop region. Re-annealing of the hairpin at forces lower than 15 pN was prevented by the annealed oligonucleotide, which worked as a kinetic block. Then, the excess of oligonucleotide was washed out of the chamber to prevent any possible interference with the secondary structure \cite{bosco2014elastic}. Force-distance curves were recorded by repeatedly moving the optical trap up and down at $\sim100$ nm/s) between a maximum and a minimum force ($\sim 1$ pN and $\sim 40$ pN), and measuring the force and the distance $\lambda$ between the trap center and the tip of the micropipette [Fig.~\ref{fig:1}(b)]. About 5-10 stretching (force increasing) and releasing (force decreasing) curves were recorded for several tethers ($\sim$ 5-10 in each condition). At low forces ($\sim 1$ pN) the ssDNA attached non-specifically to the bead surface. However, when low forces were avoided (forces above $\sim 1$ pN), no hysteresis was observed between stretching and releasing curves, showing that pulling curves were reversible.

The molecular extension of ssDNA $X_{\rm ssDNA}(f)$ is related to the extension measured experimentally, $\lambda$ [Fig.~\ref{fig:1}(a)], as follows:
\begin{equation}
X_{\rm ssDNA}(f)=\lambda(f)-x_h(f)-x_b(f)-x_{0},
\label{eq:lambda_long}
\end{equation}
where $x_h(f)=x_{h1}(f)+x_{h2}(f)$ is the extension of the two 29bp handles (characterized in \cite{forns2012handles}) $x_0$ is an arbitrary shift of the trap position (see App.~\ref{S:Ap.ssDNA_long} for details) and $x_b(f)$ is the displacement of the bead from the center of the optical trap. The latter is given by $x_b=f/k_b$ with the trap stiffness $k_b$ measured with a power spectrum analysis (See Sec. S3 in the Supplemental Material at [URL will be inserted by publisher] for details). The force plotted versus the extension $X_{\rm ssDNA}$ defines the force-extension curve (FEC).

The binding of the oligonucleotide to the loop region introduces a correction to Eq.~(\ref{eq:lambda_long}) due to the hybridized dsDNA segment. We have verified that this correction is negligible for molecules of 200b or larger (See Sec. S4 in the Supplemental Material at [URL will be inserted by publisher] for details). However, for the shortest 120b hairpin, the correction would be significant. Therefore, we used an alternative approach, the so-called two-branches method \cite{Zhang1330}.
This method implies the use of a longer loop (20b) to slow down the loop formation process, allowing measuring the ssDNA extension between $\sim$ 3-4 pN and $\sim 15$ pN (See App.~\ref{S:Ap.ssDNA_short}). The two branches method compares differences in extension between the folded and unfolded branches, without considering corrections due to the optically trapped bead and the handles. Instead, in the blocking oligonucleotide method bead and handles extensions must be subtracted, leading to higher uncertainties.

The FECs of the ssDNA for the eight different molecules studied are shown in Fig.~\ref{fig:1}(c). The FECs are scaled by dividing the measured extension over the number of nucleotides of each sequence, $x_{\textrm{ssDNA}}=X_{\textrm{ssDNA}}/N$, as shown in Fig.~\ref{fig:1}(d).

\subsection{\label{subsec:elastic_models} Elastic models}
\subsubsection{Worm-like chain}
Different polymer models have been proposed to describe the ideal elasticity (i.e. without secondary structure) of nucleic acids \cite{SMITH1996,marko1995stretching,bouchiat1999estimating}. A widely used model is the Worm-Like Chain (WLC) model with the following interpolation formula between the low- and high-force regimes \cite{marko1995stretching}:
\begin{equation}
f(x)=\frac{k_BT}{p}\left(\frac{1}{4}\left(1-\frac{x}{N\,l}\right)^{-2}-\frac{1}{4}+\frac{x}{N\,l}\right),
\label{eq:wlc_marko_siggia}
\end{equation}
where $x$ is the polymer extension at a given applied force $f$, $l$ is the contour length per base, $p$ is the persistence length and $N$ is the number of monomers of the ideal chain (with total contour length given by $L=N \cdot l$). Here we used Equation ~(\ref{eq:wlc_marko_siggia}) to reproduce the ideal ssDNA elastic behavior at 10 mM \ce{MgCl2} (Fig. \ref{fig:3}), with elastic parameters taken from Ref.~\cite{bosco2014elastic} 
($p=0.75$ nm, $l=0.69$ nm).

Previous to fitting the data to the helix-coil model the measured ssDNA extensions were corrected by $\lesssim5\%$ to match the WLC model at $f=15$ pN  (See Table S4 in the Supplemental Material at [URL will be inserted by publisher] for the used factors).

\subsubsection{Saleh formula}
\label{subsec:saleh}

Previous studies have shown that, at low forces, the measured ssDNA FECs deviate from the WLC predictions \cite{Saleh2009,McIntosh2011}. These deviations are mainly due to excluded volume effects, and they become specially relevant at low salt concentrations.
Saleh and collaborators have proposed different formulae for the regimes of low and high forces in sodium conditions \cite{Saleh2009,McIntosh2011, Jacobson2017}: at low forces the molecular extension shows a power law dependence with force due to the excluded volume effects; at high forces the elastic response is described by a WLC model with a salt-dependent internal electrostatic tension; finally, a logarithmic dependence interpolates the two force regimes (intermediate force regime). We have combined these results in a single elastic response function covering the three force regimes (low, intermediate, high). The fitting function will be denoted as Saleh-interpolating formula and is given by \cite{Saleh2009,McIntosh2011, Jacobson2017},

\begin{eqnarray}
 x(f)=\begin{cases}
    l_1\left(\frac{f}{f_1}\right)^{\gamma}, & \text{if } f\leq f_1; \\
    l_1\left(\gamma \log\left(\frac{f}{f_1}\right)+1\right), & \text{if } f_1<f< f_2 ; \\
    l_2\left(\sqrt{\frac{k_{\rm B}T}{4 l_{p}\left(f+f_{el}\right)}}\right), & \text{if } f_2\leq f . 
 \end{cases}
 \label{eq:saleh_formula}
\end{eqnarray}

with $f_1=a \left(1000c\right)^{\gamma}$, $a=0.06$ pN, $\gamma=0.054$, $l_2=0.7$ nm, $l_1=l_2/\left(2(\gamma\log(f_2/f_1)+1)\right)$ (continuity at $f_2$) and $f_2=\frac{k_{\rm B}T}{l_{p}}-f_{el}$. $c$ is the salt concentration (in M). The electrostatic tension $f_{el}$ reads:

\begin{equation}
f_{el}=\frac{k_{\rm B}T\,l_{\rm B}}{b^2}\left[\kappa b\frac{e^{-\kappa b}}{1-e^{-\kappa b}}-\log\left(1-e^{-\kappa b}\right)\right],
    \label{eq:electrostatic_tension}
\end{equation}

with an average charge length $b=1.14$ nm \cite{Saleh2009}.  $l_{\rm B}$ and $\kappa$ are the Bjerrum length and the inverse of the Debye length, respectively.

Note that the dependencies described in Eq.~\ref{eq:saleh_formula} have been only tested in sodium conditions. However, Eq.~\ref{eq:saleh_formula} can be extrapolated to magnesium by using the 100:1 salt rule for the equivalence of the screening ion effect of monovalent and divalent ions \cite{bosco2014elastic,bizarro2012non}.
In the App.~\ref{S:Ap.glyoxal} we show results of the ssDNA elasticity in presence of glyoxal (which removes secondary structure). At high salts, the WLC and the Saleh formula are compatible with the experimental data, however there are differences at low salts where the Saleh formula reproduces better the data. 
Consequently, we have used the Saleh formula (Eq. \ref{eq:saleh_formula}) to describe the ideal ssDNA elasticity at different salt conditions (Fig.~\ref{fig:4}). Comparison of the results for the  fitting parameters of the helix-coil model, $\epsilon$ and $\gamma$, using the WLC and the Saleh formula are shown in Fig.~S4 in the Supplemental Material [URL will be inserted by publisher].

The measured ssDNA extensions were corrected by $\lesssim5\%$ to match the Saleh formula model at $f=15$ pN  (See Tables~S5 and S6 in the Supplemental Material at [URL will be inserted by publisher] for the used factors).

\subsection{\label{sec:2S-model} Helix-coil model for secondary structure}

The FECs of individual ssDNA molecules deviated from the ideal polymer behavior due to the formation of secondary structures at forces below $\sim15$ pN (Figs.~\ref{fig:1}B-D). Here, we develop a helix-coil cooperative model that describes the formation of secondary structures and explains the experimental FECs within full force range under study (0-40 pN).

\begin{figure}[!ht]
\centering\includegraphics[width=\textwidth]{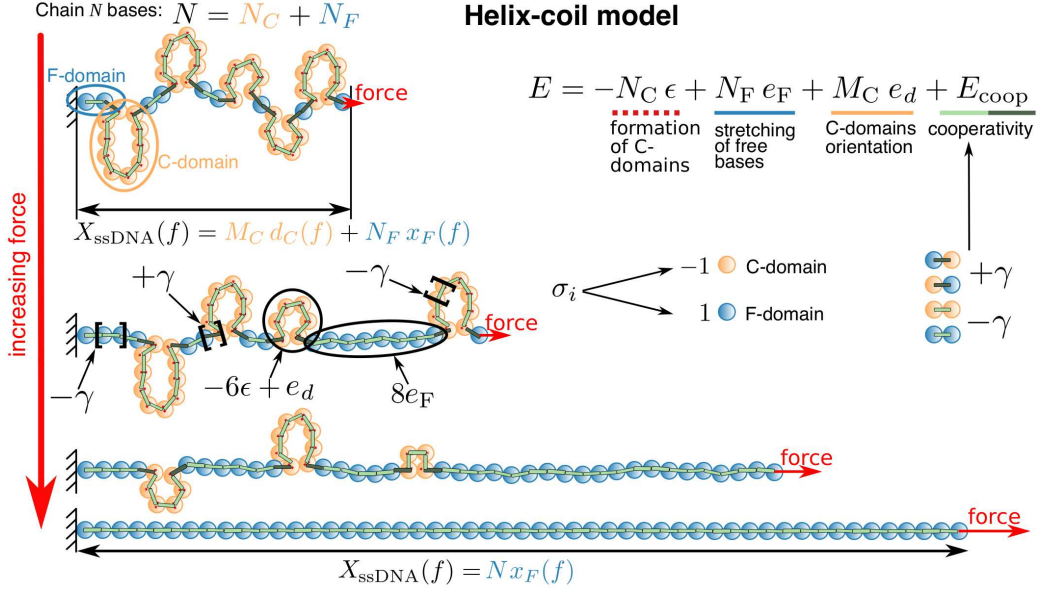}
\caption{\label{fig:2} \scriptsize{{\bf Helix-coil model for secondary structure formation.} Each monomer can be either in the free state ($\sigma_i=1$ in F-domain, blue) or in the compact state ($\sigma_i=-1$ in  C-domain, orange). Each domain type has an elastic contribution to the energy: $e_{\textrm{F}}=-\int_0^f x_{\textrm{F}}df$, per monomer in F-domain ($N_{\textrm{F}}$ being the total monomers in F-domains), and $e_{\textrm{d}}=-\int_0^f d_{\textrm{C}} df$, per C-domain ($M_{\rm C}$ being the total number of C-domains). A stretch of consecutive monomers in $\sigma_i=-1$ constitutes a C-domain with an energy per base $\epsilon$. The model also includes a cooperativity term ($E_{coop}$):
equal state adjacent monomers are energetically favored ($-\gamma$, light green) with respect to adjacent monomers in a domain wall separating C from F domains ($+\gamma$, dark green). At sufficiently low forces, the majority of the monomers are organized in C-domains. As force increases, C-domains unfold, gradually forming larger F-domains. At high forces ($\sim 15$ pN), a fully stretched unfolded state is reached, where all the monomers are in the F-domain.}}
\end{figure}

We envision the ssDNA as forming a sequence of structured compact domains interspersed by chains of free monomers (Fig.~\ref{fig:2}). These two domain types are denoted by C (compact) or F (free chain). The ssDNA is then modeled as a chain of $N$ monomers (bases) that can be in two states $\sigma_i=\pm 1$: if $\sigma_i=-1$, the monomer is part of a C-domain; otherwise ($\sigma_i=+1$), belongs to an F-domain. F-type domains pulled by a force $f$ follow the WLC model or the Saleh formula  (Sec.~\ref{subsec:elastic_models}). C-type domains are stabilized by an average energy per monomer $\epsilon$. The model includes cooperativity through an interfacial energy parameter $\gamma$ that penalizes (rewards) adjacent monomers in different (equal) domain type. Let $N_{\textrm{F}}=\sum_i^{N}\delta_{\sigma_i,1}$ and $N_{\textrm{C}}=\sum_i^{N}\delta_{\sigma_i,-1}$ be the total number of monomers of F-type and C-type, respectively, fulfilling $N=N_{\textrm{F}}+N_{\textrm{C}}$. The model Hamiltonian is given by:

\begin{eqnarray}
   \mathcal{H} = &-\epsilon\, N_{\textrm{C}}(f)-M_{\rm C}(f)\int_{0}^{f}d_{\textrm{C}}(f')df'- \nonumber \\ 
   &-N_{\textrm{F}}(f)\int_0^{f}x_{\textrm{F}}(f')df'-\gamma\sum_i^{N}\sigma_i\sigma_{i+1}.
\label{eq:hamiltonian_1}
\end{eqnarray}

The first two terms account for the total energy of the compact domains pulled at a force $f$; $M_{\rm C}(f)$ is the number of C-domains, each contributing with an extension $d_{\textrm{C}}(f)$; the third term is the stretching energy of F-type monomers with the extension per monomer $x_{\textrm{F}}(f)$, obtained by Eq.~(\ref{eq:wlc_marko_siggia}) or Eq.~(\ref{eq:saleh_formula}) (\cite{severino2019efficient}); and the last term stands for the interfacial energy between adjacent monomers. A schematic representation of this model is shown in Fig.~\ref{fig:2}. At a given force, $N_{\textrm{F}}$ monomers with $\sigma=+1$ contribute by $x_{\textrm{F}}(f)$ to the total extension. The rest of monomers, $N_{\textrm{C}}$, are distributed into $M_{\rm C}$ compact domains, each one contributing by $d_{\textrm{C}}(f)$ to the full extension. Hence, the total molecular extension is given by: $X_{\rm ssDNA}(f)=M_{\rm C}(f) d_{\textrm{C}}(f)+N_{\textrm{F}}(f) x_{\textrm{F}}(f)$.

The Hamiltonian in Eq.~(\ref{eq:hamiltonian_1}) is analytically solvable (See Sec.~\ref{sec:2S-model}), leading to the extension per monomer, $X_{\textrm{ssDNA}}/N$, 
 \begin{equation}
 x_{\rm ssDNA}(f)=\phi_{\textrm{F}}(f)x_{\textrm{F}}(f)+\frac{M_{\rm C}(f)}{N}d_{\textrm{C}}(f),
  \label{eq:extension_model_old}
 \end{equation}
 where the fraction of free monomers, $\phi_{\textrm{F}}(f)=N_{\textrm{F}}(f)/N$, is given by \footnote{$\phi_{\textrm{F}}$ equals the fraction of unpaired bases $\phi$ defined in Ref.~\cite{bosco2014elastic}.}:
\begin{equation}
\phi_{\textrm{F}}(f)=\frac{1}{2}\left(1+\frac{\sinh\left(\beta A(f)\right)}{\sqrt{e^{-4\beta B(f)}+\sinh^2\left(\beta A(f)\right)}}\right),
\label{eq:fraction_ssDNA}
\end{equation}

with $\beta=1/k_{\rm B}\,T$. A and B quantities are defined by

\begin{eqnarray}
   A(f) & = & -\frac{\epsilon}{2}+\frac{1}{2} \int_0^{f}x_{\textrm{F}}(f')df'\,\,;  \\ 
   B(f) & = & -\frac{1}{4}\int_0^{f}d_{\textrm{C}}(f')df'+\gamma.
\end{eqnarray}
Eq.~(\ref{eq:fraction_ssDNA}) can be used to obtain the fraction of bases in C-domains, $\phi_{\textrm{C}}(f)=N_{\textrm{C}}(f)/N$, as
\begin{equation}
\phi_{\textrm{C}}(f)=1-\phi_{\textrm{F}}(f). 
\label{eq:fraction_compact}
\end{equation}
The number of C-domains is given by:
\begin{equation}
M_{\rm C}(f)=\frac{N}{2}\frac{\sinh\left(\beta A(f)\right)}{\cosh\left(\beta A(f)\right)+\sqrt{e^{-4\beta B(f)}+\sinh^2\left(\beta A(f)\right)}}.
\label{eq:number_hairpins}
\end{equation}
The average number of monomers per C-domain reads as:
\begin{equation}
n_{\textrm{C}}(f)=\frac{N\phi_{\textrm{C}}(f)}{M_{\rm C}(f)}.
\label{eq:nav}
\end{equation}

In Eq.~(\ref{eq:extension_model_old}), $x_{\textrm{F}}(f)$ is given by the WLC model [Eq.~(\ref{eq:wlc_marko_siggia})] or the Saleh formula [Eq.~(\ref{eq:saleh_formula})], with the parameters described in Sec.~\ref{subsec:elastic_models}. For the extension $d_{\textrm{C}}(f)$,  we consider each domain as a rigid object of length $\delta$ that aligns along the stretching direction (App.~\ref{S:Ap.hairpin_orient}). For example, a hairpin-like C-domain initiating with a stem would have a length equal to the diameter of a B-DNA double-helix, $\delta\sim 2$ nm \cite{saengerprinciples2013}). We checked that this parameter does not affect the predictions of the model significantly (App.~\ref{S:Ap.hairpin_orient}). Therefore, we neglected the domain orientation term ($\delta=0$) in the analysis and fitted our data to the simpler expression:

\begin{equation}
 x_{\rm ssDNA}(f)=\phi_{\textrm{F}}(f)x_{\textrm{F}}(f).
  \label{eq:extension_model}
 \end{equation}

We fitted the model [Eq.~(\ref{eq:extension_model})] to the experimental FECs  with $\epsilon$ and $\gamma$ as fitting parameters between $2 \leq f\leq 15$ pN. Lower forces ($f<2$ pN) were not considered in the fit to avoid any influence of incorrect tethering (specially in molecules with $N\lesssim1000$b). Importantly, the model reproduces the experimental FECs throughout the whole force range $0 < f\leq 15$ pN  in the different conditions tested (different length and salts).

 \begin{figure}[!ht]
\centering\includegraphics[width=\textwidth]{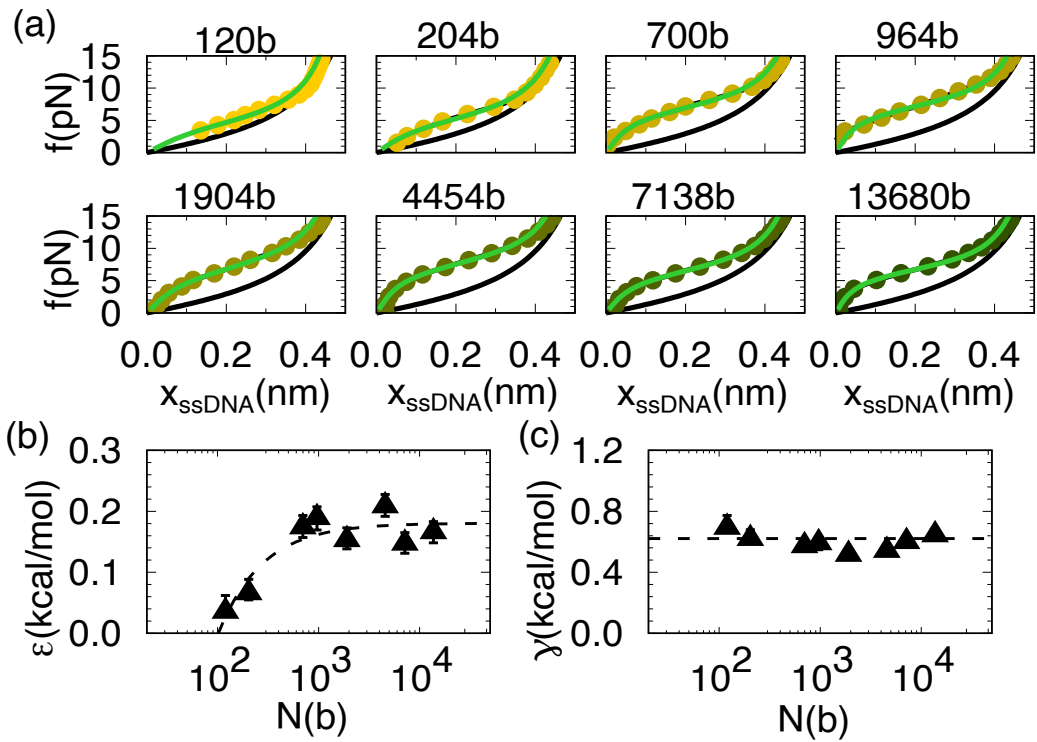}
\caption{\label{fig:3} \scriptsize{{\bf ssDNA FECs for varying molecular length.}
(a) Experimental FECs (circles) and the corresponding
fits of the helix-coil model [Eq.~(\ref{eq:extension_model})] (green lines) for ssDNA molecules with lengths ranging from $\sim$120b to $\sim$14000b [color code as in Fig.\ref{fig:1}(c)-(d)]. The WLC model [Eq.~(\ref{eq:wlc_marko_siggia})] is shown as a black line. Fits give the energy parameters $\epsilon$ and $\gamma$ for each molecular length.
(b,c) Average formation energy per base in C-domain, $\epsilon$ (b), and cooperativity parameter, $\gamma$ (c), obtained from the fits shown in panel (a) as a function of the ssDNA molecular length (triangles). Dashed lines are fits to Eq.~(\ref{eq:energy_frac_length}) ($\epsilon_0=0.18(1)$ kcal/mol and $b=18(4)$ kcal/mol) and to a constant value ($\gamma=0.61(2)$ kcal/mol). Experiments were performed at \SI{25}{\celsius} in 10 mM \ce{MgCl2}. Error bars are the statistical errors.}}
\end{figure}

\begin{figure}[!ht]
\centering\includegraphics[width=\textwidth]{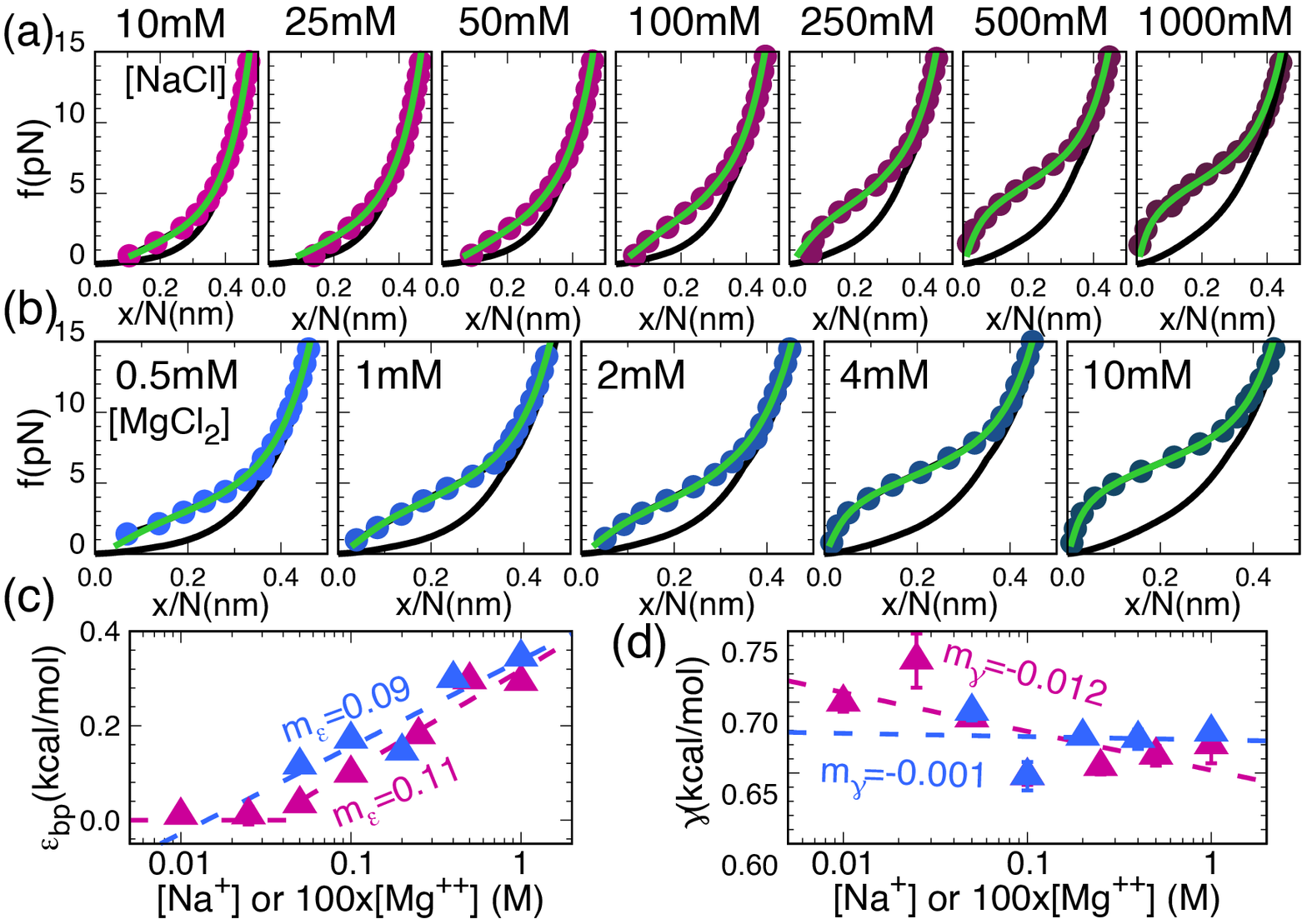}
\caption{\label{fig:4} \scriptsize{{\bf ssDNA FECs for varying salt conditions.}
(a,b) Fits of the helix-coil model [Eq.~(\ref{eq:extension_model})] (green lines) to the FECs of 13680b ssDNA at increasing NaCl  concentration (magenta circles, a) and MgCl$_2$ concentration (blue circles, b). The Saleh formula [Eq.~(\ref{eq:saleh_formula})] is shown as a black line. 
Fits give the energy parameters $\epsilon$ and $\gamma$ for each salt condition. (c,d) Dependence of $\epsilon_{\rm bp}$ ($\epsilon_{\rm{bp}}=2\epsilon$) and $\gamma$ with salt concentration (magenta and blue triangles for sodium and magnesium, respectively). Data in (c) above 50mM were fitted to Eq.~(\ref{eq:energy_salt_conc}) (dashed lines) giving $\epsilon^{0}_{\rm{bp}}=0.34(2)$ kcal/mol and $m_{\epsilon}=0.11(2)$ kcal/mol for \ce{NaCl} (magenta) and $\epsilon^{0}_{\rm{bp}}=0.75(9)$ kcal/mol and $m_{\epsilon}=0.09(2)$ kcal/mol for \ce{MgCl2} (blue). The values for $\epsilon$ below $[\ce{NaCl}]\sim 50$mM appear to saturate to 0. $\gamma$ also follows a logarithmic dependence with salt concentration [Eq.~(\ref{eq:gammasalt})], with $\gamma_0=0.64(1)$ kcal/mol and $m_{\gamma}=-0.012(5$ kcal/mol for \ce{NaCl} and $\gamma_0=0.67(2)$ kcal/mol and $m_{\gamma}=-0.001(4)$ for \ce{MgCl2}. Error bars are the statistical errors.}}

\end{figure}
 
\section{Results}
\subsection{Force-extension curves}

For ssDNA measurements a DNA hairpin was fully unzipped by applying a force above 15 pN at the 3' and 5' extremities of the molecule, exposing the loop region. Unzipping of the hairpin was performed in the presence of an oligonucleotide complementary to the loop (Sec.~\ref{subsec:molecular_extension}). 
Fig.~\ref{fig:1}(b) shows the pulling curve (red) corresponding to the unzipping of the H$_{7138}$ hairpin with the characteristic sawtooth pattern \cite{Huguet2010PNAS}. This is followed by the relaxation curve (grey), after the oligonucleotide hybridizes to the loop region, corresponding to the ssDNA FEC. It differs from the ideal elastic curve (WLC model shown in black) at low forces, where ssDNA compacts and the extension shortens showing the characteristic shoulder of secondary structure formation \cite{bosco2014elastic,Maier2000}.  It  is  well  known  that DNA sequence (i.e. GC content) and monovalent and divalent cations (which screen the negatively charged DNA phosphate backbone)  affect secondary structure formation \cite{Croquette2001,bosco2014elastic,Bochman2012}. However, the effect of ssDNA length has never been studied. 
The nature of the compact structure is an open question due to its highly disordered and dynamic features. A dependence of the FEC with the ssDNA length might be the signature of long-range interactions along the chain, giving information about the nature of the secondary structure. 

\subsection{\label{subsec:length_effects}Effects of molecular length on secondary structure}
We measured the FECs of eight different ssDNA molecules [Fig.~\ref{fig:1}(c)] with lengths spanning from 120 to 13680b in 10 mM \ce{MgCl2} (Sec.~\ref{subsec:buffers}) and similar GC content, $\sim 50\%$ \footnote{A mild trend with varying GC content has been observed, see App.~\ref{S:Ap.GC}.}. 
FECs are shown in Fig.~\ref{fig:1}(d) with the extension rescaled by the number of ss bases of each molecule. Molecules with $N\gtrsim500$b approximately collapse into a single curve, deviating from the ideal elastic behavior at forces $f\lesssim 10$ pN and showing a remarkable shoulder. Small differences are due to different GC content (from 44\% to 53\%) which correlates with the height of the force shoulder observed in the FECs (see below and App.~\ref{S:Ap.GC}).
In contrast, for $N\lesssim500$b FECs do not collapse into the same curve and the height of the shoulder decreases with $N$, indicating finite-length effects. This might be due to the depletion of secondary structure arrangements in the regime $N\lesssim500$b.

As shown in Fig.~\ref{fig:3}(a), the FECS for the studied molecules (yellow to dark green circles for increasing lengths) are well reproduced by Eq.~(\ref{eq:extension_model}) in the proposed model (green curves). The best-fitting values for the average energy per base of C-domains, $\epsilon$, and the cooperativity parameter, $\gamma$, as a function of the molecular length are shown in Figs.~\ref{fig:3}(b)-(c).
 $\epsilon$ is found to follow a phenomenological $1/N$-correction, 
\begin{equation}
    \epsilon(N)=\epsilon_0-\frac{b}{N},
    \label{eq:energy_frac_length}
\end{equation}
with $\epsilon_0=0.18(1)$ kcal/mol and $b=18(4)$ kcal/mol [Fig.~\ref{fig:3}(b)]. $\gamma$ appears to be constant [$\gamma=0.61(2)$ kcal/mol] for all ssDNA lengths [Fig.~\ref{fig:3}(c)]. Equation~(\ref{eq:energy_frac_length}) quantifies corrections to the  stabilization energy of C-domains due to the free ends of the polymer. In generic statistical physics models, corrections to the equilibrium free energy are determined by the surface to volume ratios, being on the order of $1/N$ for the one-dimensional helix-coil model.

For shorter molecules ($N\lesssim500$b),  $\epsilon$ starts deviating from the saturation value $\epsilon_0$. We stress that the FECs in Fig.\ref{fig:3}(a) cannot be fitted to the model with $\gamma=0$, as shown in Fig.~S5 in the Supplemental Material at [URL will be inserted by publisher]. Therefore, cooperativity is key to reproduce the experimental results, suggesting the suitability of the helix-coil model.

\subsection{\label{subsec:salt_effects} Salt dependence of secondary structure}
In order to investigate how monovalent and divalent ions affect ssDNA folding, we fitted the helix-coil model to previously published data by us of a ssDNA molecule (H$_{13680}$, $\sim50$\% GC content) under different \ce{NaCl} and \ce{MgCl2} concentrations \cite{bosco2014elastic}. FECs for H$_{13680}$ are shown in Fig.~\ref{fig:4}(a) (\ce{NaCl}) and Fig.~\ref{fig:4}(b) (\ce{MgCl2}).
At high salts ($\gtrsim$ 100 mM \ce{NaCl}, $\gtrsim$ 0.5 mM \ce{MgCl2}), the data (circles) deviate from the ideal elastic behavior (black curves). The height of the shoulder in the FEC reaches its maximum at the highest salt concentrations ([\ce{NaCl}]=1000 mM and [\ce{MgCl2}]=10 mM). At low salts ($\lesssim$ 50mM \ce{NaCl}), the shoulder disappears and the FECs approach the ideal elastic behavior \footnote{At low forces the measured extension is slightly larger than predicted by the WLC model, suggesting excluded volume effects \cite{bosco2014elastic,marenduzzo2003485,toan200580,toan2006}.}.

The helix-coil model successfully reproduces the FECs in sodium and magnesium [Figs.~\ref{fig:4}(a)-(b)]. For high salts (above 100mM \ce{NaCl} and 1mM \ce{MgCl2} the WLC model Eq.~(\ref{eq:wlc_marko_siggia}) and the Saleh formula Eq.~(\ref{eq:saleh_formula})  give compatible results. Otherwise, at low salts, the Saleh formula reproduces better the ideal elastic response of ssDNA (App.~\ref{S:Ap.glyoxal}) so we used it throughout the following figures. A comparison of the two models (WLC versus Saleh) for the energy parameters $\epsilon$, $\gamma$ and the FECs is shown in Figs.~S4 and S6 in the Supplemental Material at [URL will be inserted by publisher]. Salt-induced stabilization of secondary structure is captured by the model as an increase in $\epsilon$ with \ce{NaCl} (magenta) and \ce{MgCl2} (blue) concentrations. Note that, for $\epsilon=0$ and zero force, F- and C-domains are equally probable meaning that half of the bases are of C- and F-type. $\epsilon$ values can be compared with the Watson-Crick (WC) base-pair energies of the nearest-neighbour (NN) model. Therefore, we define $\epsilon_{\rm{bp}}=2\epsilon$ as the equivalent of the NN base-pair energy for DNA hybridization. Similarly to the energies in the NN model \cite{Huguet2010PNAS,SantaLucia1998}, we find that $\epsilon_{\rm{bp}}$ depends logarithmically with salt concentration [Fig.~\ref{fig:4}(c)]: 

\begin{equation}
    \epsilon_{\textrm{bp}}(\textrm{c})=\epsilon_{\textrm{bp}}^0+m_{\epsilon}\log\left(\textrm{c}\right),
    \label{eq:energy_salt_conc}
\end{equation}

\begin{table}
\begin{tabular}[width=\textwidth]{cccc}
Condition & Relation & & Values (kcal/mol) \\
\hline
\small{Length}  & \small{$\epsilon=\epsilon_0+b/N$} & & \small{$\epsilon_0= 0.18(1)$, $b=18(4)$}\\
\small{(10mM \ce{MgCl2})} & \small{$\gamma=$constant} & & \small{$\gamma=0.61(2)$} \\
\hline
 & & \small{c=[NaCl]} & \small{$\epsilon_{\textrm{bp}}^0=0.34(3)$, $m_{\epsilon}=0.11(2)$}  \\
 \small{Salt} &  \small{$\epsilon_{\textrm{bp}}(\textrm{c})=\epsilon_{\textrm{bp}}^0+m_{\epsilon}\log\left(\textrm{c}\right)$} & & \small{$\gamma_0=0.64(2)$, $m_{\gamma}=-0.012(5)$} \\
 & \small{$\gamma(c)=\gamma_0+m_{\gamma}\log\left(\textrm{c}\right)$} & c=[\ce{MgCl2}] & \small{$\epsilon_{\textrm{bp}}^0=0.75(9)$, $m_{\epsilon}=0.09(2)$}  \\
& & & \small{$\gamma_0=0.67(2)$, $m_{\gamma}=-0.001(4)$} \\
\end{tabular}
\caption{\label{tab:table1}
Fitting parameters obtained for each studied condition.}
\end{table}

with $\rm{c}$ being the salt concentration (in M units), $\epsilon_{\rm {bp}}^0$ the value at 1M and $m_{\epsilon}$ the salt correction parameter for secondary structure. Equation~(\ref{eq:energy_salt_conc}) is a phenomenological expression that follows from applying standard physical chemistry theories of thermodynamic activity to diluted ionic solutions. Fits of $\epsilon_{\rm {bp}}$ to Eq.~(\ref{eq:energy_salt_conc}) are shown in Fig.~\ref{fig:4}(c) as dashed lines. The results from the fits are $\epsilon_{\rm {bp}}^0=0.34(3)$ kcal/mol, $m_{\epsilon}=0.11(2)$ kcal/mol in sodium and $\epsilon_{\rm {bp}}^0=0.75(9)$ kcal/mol, $m_{\epsilon}=0.09(2)$ kcal/mol in magnesium. The values for $\epsilon_{\rm {bp}}^0$ are $\sim 5$ times smaller than the average NN base-pair energy ($\sim 1.6$ kcal/mol) and almost 3 times smaller than the most unstable NN bp (AT/TA$\sim0.85$ kcal/mol at 1 M \ce{NaCl} and 10 mM \ce{MgCl2}) \cite{Huguet2010PNAS,SantaLucia1998}. 
Such a lower stabilization is expected for random sequences that lack full complementarity regions. In fact, C-domains might consist of aggregates of base pairs, but also large loops
and mismatches, which decrease base-pairing stability \cite{hooyberghs2009effects,landuzzi2020mismatch}. In agreement with this, unzipping forces in dsDNA are $\sim15$ pN [Fig.~\ref{fig:1}(b)] (see also \cite{Woodside2006science}), much larger than the force at the shoulder in the FEC ($\sim5$-$10$ pN).
Interestingly, the slope $m_{\epsilon}$=0.11(2) kcal/mol [Eq.~(\ref{eq:energy_salt_conc})] in sodium is compatible with the homogeneous salt correction parameter of the unified oligonucleotide dataset \cite{Peyret2000} and with that derived from unzipping experiments \cite{Huguet2010PNAS} ($m=0.114$ kcal/mol and $m=0.104$ kcal/mol, respectively).
For the \ce{MgCl2} case, $m_{\epsilon}=0.09(2)$ kcal/mol is about twice the Mfold value \cite{Zuker2003} ($m=0.055$ kcal/mol)  but closer to the unzipping value \cite{Huguet2017} [$m=0.07(2)$ kcal/mol]. Note that Mfold assumes that the salt correction for sodium is exactly twice that of magnesium ($m^{\ce{NaCl}}_{\epsilon}=2\cdot m^{\ce{MgCl2}}_{\epsilon}$). 

Finally, for the cooperativity term $\gamma$, we have also assumed the phenomenological logarithmic salt dependence of Eq.(\ref{eq:energy_frac_length}) [Fig.~\ref{fig:4}(d)]:
\begin{equation}
\gamma=\gamma_0+m_{\gamma}\log\left(\textrm{c}\right), 
\label{eq:gammasalt}
\end{equation}
with $\rm{c}$ being the salt concentration (in M units), $\gamma_0$ the value at 1M and $m_{\epsilon}$ the salt correction parameter. We get $\gamma_0=0.64(2)$ kcal/mol and  $m_{\gamma}=-0.012(5)$ kcal/mol for \ce{NaCl} and $\gamma_0=0.67(2)$ kcal/mol and $m_{\gamma}=-0.001(4)$ kcal/mol for \ce{MgCl2}.
Although $\gamma$ weakly depends on $c$ ($m_{\gamma}<0$) for sodium, the dependence is negligible for magnesium. This indicates that salt screening effects for the cooperativity between adjacent bases are weak raising the question whether it depends at all. The salt dependence of $\gamma$ [Fig.~\ref{fig:4}(d)] is the same if we apply the 100th rule of thumb between sodium and magnesium concentrations, as we did for $\epsilon_{\rm bp}$ [Fig.~\ref{fig:4}(c)]. The values of $\gamma$ at the reference salt condition, $\gamma_0$, are compatible with the average $\gamma$ obtained with varying molecular length at the same condition [$\gamma=0.61(2)$ kcal/mol, see Sec.~\ref{subsec:length_effects}].  
Notice that the switching of a base (from F to C or the opposite) involves an interfacial energy cost/gain of  $\pm2\gamma$ [Eq.(\ref{eq:hamiltonian_1})]. Note the opposite sign of the salt corrections for $\epsilon$ and $\gamma$. In both cases, salt screens the electrostatic interactions by weakening phosphates repulsion. This leads to an increased base pairing stabilization and a (much weaker) decreased cooperativity.

A compendium of all results in this section is shown in Table~\ref{tab:table1} and the fitting parameters obtained with the WLC model are shown in Table~S7 in the Supplemental Material at [URL will be inserted by publisher].

\subsection{C- and F-domains organization}

\begin{figure}[!ht]
\centering\includegraphics[width=0.9\textwidth]{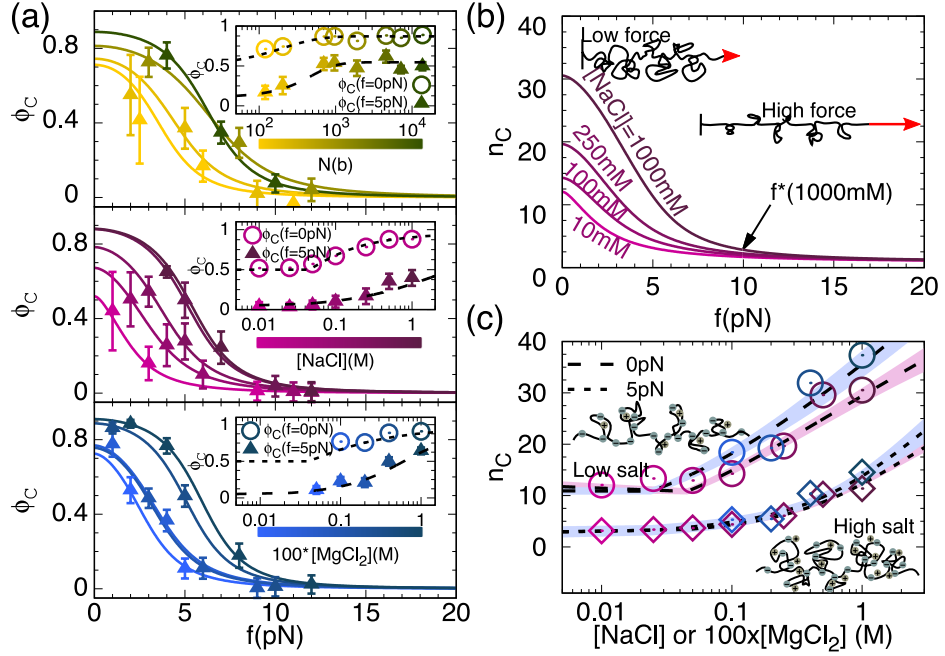}
\caption{\label{fig:5} \scriptsize{{\bf C- and F-domains organization.} (a) Fraction of bases in C-domains, $\phi_{\textrm{C}}$, as a function of force, obtained from Eq.~(\ref{eq:fraction_compact}) using the best fitting values for $\epsilon$ and $\gamma$  [Fig.~\ref{fig:3}(b)-(c) and Fig.~\ref{fig:4}(c)-(d)] for: (top) four different molecular lengths (120, 204, 1904 and 13680b, from yellow to dark green), (middle) five NaCl concentrations (10, 100, 250, 500 and 1000 mM, from light to dark magenta) and (bottom) five \ce{MgCl2} concentrations (0.5, 1, 2, 4 and 10 mM, from light to dark blue). The insets show the dependence of the fraction $\phi_{\textrm{C}}$ with respect to the molecular length (top) and salt concentration (middle and bottom) at two different forces, 0 and 5 pN. 
Dashed lines are the theoretical prediction by Eqs.~(\ref{eq:fraction_ssDNA}) and (\ref{eq:fraction_compact}) using a salt and length dependent $\epsilon$ and $\gamma$ parameters as given in Table~\ref{tab:table1}. (b) Average number of bases or size per C-domain, $n_{\textrm{C}}$ [Eq.~(\ref{eq:nav}), with salt dependent $\epsilon$ and $\gamma$ parameters given in Table~\ref{tab:table1}], as a function of force for 10, 100, 250 and 1000mM NaCl. The schematic depictions show a highly compacted form of the ssDNA at low forces that is stretched at higher forces, decreasing the size of the C-domains. The arrow indicates the threshold force $f^*$. (c) $n_{\textrm{C}}$ dependence on salt concentration (magenta for NaCl and blue for \ce{MgCl2}) at 0 and 5 pN. Symbols correspond to Eq.~(\ref{eq:nav}) using the best fitting values for $\epsilon$ and $\gamma$ [Fig.~\ref{fig:4}(c)-(d)]. Dashed lines are the theoretical prediction by Eq.~(\ref{eq:nav}) with salt dependent $\epsilon$ and $\gamma$ parameters given in Table~\ref{tab:table1}. Shadowed areas show the uncertainty region of $n_{\textrm{C}}$, from the errors in $\epsilon$ and $\gamma$. The schematic depictions show the negatively charged ssDNA (black line) screened by the cations (grey dots) in the solvent stabilizing C-domains.}}

\end{figure}

We use the proposed helix-coil cooperative model (Sec.~\ref{sec:2S-model}) to investigate how bases are distributed between F- and C-domains. To this end, we compute the fraction of bases belonging to C-domains at each force $f$, $\phi_{\textrm{C}}(f)$ [Eq.~(\ref{eq:fraction_compact})], and the average number of bases per C-domain, $n_{\textrm{C}}(f)$ [Eq.~(\ref{eq:nav})]. 

Fig.~\ref{fig:5}(a) shows $\phi_{\textrm{C}}(f)$ as a function of force at different ssDNA lengths (top), \ce{NaCl} (middle) and \ce{MgCl2} (bottom) concentrations obtained from fitting the model to the experimental data. Insets in Fig.~\ref{fig:5}(a) show $\phi_{\textrm{C}}$ for each condition at two different forces, $f=0$ and $5$ pN. $\phi_{\textrm{C}}(f)$ presents its maximum value at 0 pN, $\phi_{\textrm{C}}(0)=0.5$-$0.9$, depending on salt and length. $\phi_{\textrm{C}}(f)$ decreases until a threshold force is reached, $f^*$. Above $f^*$ the secondary structure does not form ($\phi_{\textrm{C}}(f^*)\sim 0$) and the experimental FEC matches the ideal elastic model prediction Eqs.~(\ref{eq:wlc_marko_siggia}-\ref{eq:saleh_formula}). $f^*$ depends on molecular length and salt concentration: for long molecules ($N\geq500$b) and high salts ([\ce{NaCl}]=$1000$ mM, [\ce{MgCl2}]=$10$ mM), $f^* \sim 10$ pN, whereas for either short molecules or low salts ([\ce{NaCl}]=$10-50$ mM, [\ce{MgCl2}]=$0.5$ mM), $f^*\sim 5$-$7$ pN. Note that $\phi_C(0)$ is non-negligible, even at the lowest NaCl concentrations (10-50 mM NaCl, dashed lines in the insets of Fig. 5(a), middle). In spite of this, a shoulder in the FEC is not observed [Fig.~\ref{fig:4}(a)], because $\epsilon$ is too small.

\begin{figure}[!ht]
\centering\includegraphics[width = 0.9\textwidth]{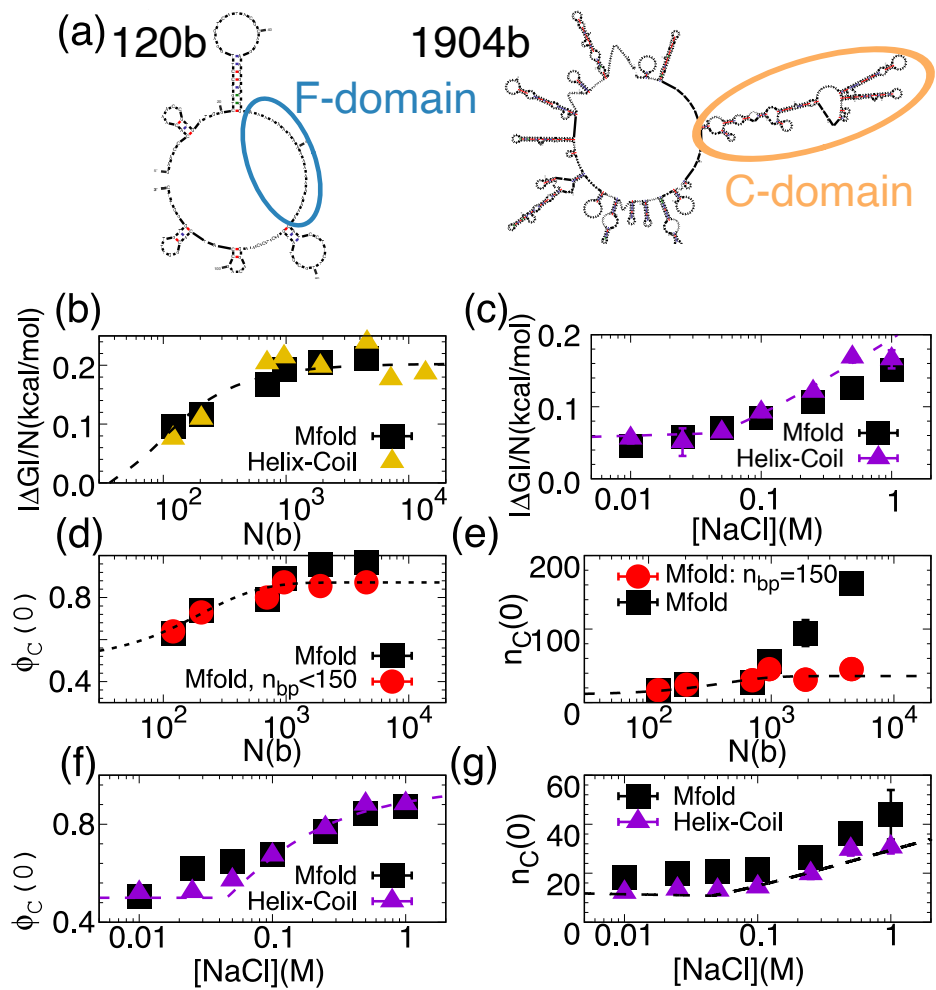}
\caption{\label{fig:6} \scriptsize{{\bf Comparison between the helix-coil and Mfold models for ssDNA secondary structure formation.}
(a) The first excited secondary structures above the native hairpin as given by Mfold for a short (120b) and a long molecule (1904b) at 10mM \ce{MgCl2}. (b) Absolute value of the free energies per base at zero force, $|\Delta G|/N$, given by Mfold (black squares) and predicted by the helix-coil model [Eq.~(\ref{eq:free_energy_model})] (dashed line) as a function of molecular length $N$.(c) Same as in (b) a function of the sodium concentration. The structures calculated for Mfold are from a 500bp hairpin (${\rm H}_{1004}$) with a random sequence with the same GC content as in ${\rm H}_{13680}$. Purple triangles correspond to the values obtained from the fits to the experimental FECs. (d,e) Fraction of bases in C-domains (d) and average number of bases $n_{\textrm{C}}$ per C-domain (e) given by Mfold (red circles and black squares) and the helix-coil model (dashed line). Red circles are obtained from Mfold by constraining the maximum distance between base pairing to 150b. Errors are smaller than the size of the symbols. (f,g) Same as in (d,e) as a function of the sodium concentration. As in (c), the structures of Mfold have been obtained for ${\rm H}_{1004}$. The agreement between Mfold and the helix-coil model predictions is in general very good.}}
\end{figure}

Fig.~\ref{fig:5}(b) shows $n_{\textrm{C}}(f)$ versus force obtained by fitting the model to the experimental data, and a similar trend to $\phi_{\textrm{C}}(f)$ is observed: it is maximum at zero force, $n_{\textrm{C}}(0)$, and decreases as force increases, until the threshold force is reached, $f^*$, for which $n_{\textrm{C}}\to 1$ (the minimum value allowed by the model). The values of $f^*$ for $\phi_{\textrm{C}}(f)$ and $n_{\textrm{C}}(f)$ are the same, indicating that secondary structure is disrupted above $f^*$. As shown in Fig.~\ref{fig:5}(c), $n_{\textrm{C}}$ increases with sodium and magnesium for all forces. The values for $n_{\textrm{C}}$ range from $\sim10$b (low salt) to $\sim30$b (high salt). Overall, the model predicts that the applied force destabilizes the secondary structure, by reducing both the size and number of C-domains [sketch in Fig.~\ref{fig:5}(b)]. 

\section{\label{sec:mfold} Comparison of the model with Mfold predictions}

Mfold predicts secondary structures of nucleic acids from their primary sequence. In order to validate our model, we compare its predictions at zero force with those obtained from Mfold for $\textrm{H}_{120}$, $\textrm{H}_{204}$, $\textrm{H}_{700}$, $\textrm{H}_{964}$, $\textrm{H}_{1904}$ and $\textrm{H}_{4454}$ \footnote{The 7138 and 13680 bases sequences were not studied due to the huge computational time (which grows like $\sim N^3$).}. 
For a given ssDNA sequence, Mfold computes folded structures of free-energy $\Delta G$ (with respect to the random coil) in a given range above the absolute minimum, defining the energy spectrum of that sequence. We analyzed each folded  structure in the spectrum [Fig.~\ref{fig:6}(a)], and assigned nucleotides as being of F- or C-type depending on whether they belong to a folded hairpin-like structure (C-domain) or they are part of a free linker (F-domain). This permits us to calculate $\phi_{\textrm{C}}$, $n_{\textrm{C}}$, by counting the number of bases forming C-domains ($\phi_{\textrm{C}}$) and the number of bases per C-domain ($n_{\textrm{C}}$). 

To compare our model predictions with Mfold, we derive an analytical expression for the free energy of the folded ssDNA at zero force from  the helix-coil model (App.~\ref{S:Ap.2SModel}):

\begin{eqnarray}
 \frac{\Delta G}{N}=-\frac{\epsilon}{2}-\frac{1}{\beta}\log\Bigl[\cosh\left(\frac{\beta\epsilon}{2}\right)+ \nonumber\\
 \sqrt{e^{-4\beta\gamma}+\sinh^{2}\left(\frac{\beta\epsilon}{2}\right)}\Bigr],
 \label{eq:free_energy_model}
\end{eqnarray}

Eq.~(\ref{eq:free_energy_model}) nicely matches the Mfold prediction with varying $N$ [Fig.~\ref{fig:6}(b)]  and varying sodium concentration Fig.~\ref{fig:6}(c)]. Note that there is no fitting procedure, we just used the values $\epsilon$ and $\gamma$ from the FEC fits (purple triangles) and the fitted phenomenological formulae [dashed lines in Fig.~\ref{fig:3}(c)-(d)] and Fig.~\ref{fig:4}(c)-(d)]. The  Mfold prediction in Fig.~\ref{fig:6}(c) corresponds to a 1004b random sequence with the same GC content as in $H_{13680}$, since the Mfold analysis cannot be performed with very long sequences (larger than $\sim$5000b, [72]). The same results are obtained for different sequence lengths whenever the GC content is conserved. It is important to stress the strong dependence of the free energy $\Delta G$ with the GC content (see Fig. S7 in the Supplemental Material at [URL will be inserted by publisher]), which must be taken into account for a proper comparison. The match between the Mfold prediction and Eq.~(\ref{eq:free_energy_model}) works better when using the Saleh formula [Eq.~(\ref{eq:saleh_formula})] as compared to the WLC [Eq.~(\ref{eq:wlc_marko_siggia})] for the ideal ssDNA elasticity  (see Fig.~S7 in the Supplemental Material at [URL will be inserted by publisher]). The equivalent of Fig.~\ref{fig:6}(c) in magnesium is not predictive because Mfold uses a salt correction that is known to be inaccurate \cite{Huguet2017}.

In addition, for short ssDNA molecules ($N<1000$b), the helix-coil model and Mfold also predict similar values for $\phi_{\textrm{C}}(0)$ and $n_{\textrm{C}}(0)$  [Fig.~\ref{fig:6}(d)-(e)]. However, for longer ssDNA molecules discrepancies emerge. While Mfold predicts that $\phi_{\textrm{C}}(0)\to 1$ [Fig.~\ref{fig:6}(d), black squares] and $n_{\textrm{C}}(0)$ rapidly increases with $N$ [Fig.~\ref{fig:6}(e), black squares], our model predicts that $\sim 10\%$ of all bases are free [Fig.~\ref{fig:6}(d), dashed line] and $n_{\textrm{C}}(0)$ saturates at $\sim 40$b [Fig.~\ref{fig:6}(e), dashed line]. These discrepancies might be attributed to the simplicity of the helix-coil model, which does not consider sequence heterogeneity (energy parameters $\epsilon$ and $\gamma$ are taken as uniform). In contrast, Mfold incorporates sequence effects favoring the hybridization of nearly complementary regions, even if far apart, a feature absent in our model. This leads to large C-domain sizes in Mfold [e.g. Fig.~\ref{fig:6}(a), right structure]. To test this, we have computed low-lying energy structures with Mfold by constraining the maximum distance between base pairs. For 150b distance, the values for $\phi_{\textrm{C}}(0)$ and $n_{\textrm{C}}(0)$ predicted by Mfold for long molecules ($N>1000$b) agree with our model [red circles in Figs.~\ref{fig:6}(d)-(e)], supporting this interpretation. As shown in Fig.~\ref{fig:6}(f)-(g), the helix-coil model also reproduces the Mfold predictions for $\phi_{\textrm{C}}(0)$ and $n_{\textrm{C}}(0)$ as a function of the salt concentration for short enough molecules [Mfold predictions are obtained for the same 1004b sequence as in Fig.~\ref{fig:6}(c)].

It is remarkable that, despite its simplicity, the helix-coil model compares so well with Mfold. Mfold is a predictive tool that uses the NN model with energy parameters derived from large datasets obtained from DNA melting experiments. The excellent agreement between our model and Mfold reflects the fact that both build upon the energetics of hybridization. The relevant motifs in Mfold (base pairs, loops, single-stranded segments, etc.) are also the ultimate building blocks of the compact disordered structures. Interestingly, the combined effects of these blocks are effectively described by only two parameters ($\epsilon$ and $\gamma$) in the helix-coil model.

\section{Discussion and Conclusion}

We have developed a cooperativity-dependent folding model that reproduces experimental FECs of ssDNA over different molecular lengths and salt conditions. 
The model assumes that ssDNA folds forming compact domains (C) interspersed by free bases (F). While free bases are described by the ideal elastic model, C-domains are assumed to have a negligible extension. Compared to more complex theoretical models \cite{hairpinssecstruc2001,SEOL2007}, the model presented here includes only two parameters: $\epsilon$, the average energy per base in a C-domain, and $\gamma$ the interfacial energy between adjacent domains (cooperativity term). With only two parameters, the model reproduces the experimental FECs of ssDNA spanning two decades of contour length ($100\,\rm{b}\lesssim N\lesssim 14$ kb) and three decades of sodium and magnesium concentrations ($10 \leq \rm{[\ce{NaCl}]}\leq 1000$ mM and $0.5 \leq \rm{[\ce{MgCl2}]}\leq 10$ mM). 
Remarkably, $\gamma=0$ is unable to fit the data suggesting that cooperativity is essential for ssDNA folding, as shown in Fig.~S5 in the Supplemental Material at [URL will be inserted by publisher]. The simplest interpretation of the cooperativity term is the nature of the stacking interaction itself. In nucleic acids two distinct forces stabilize a duplex: hydrogen bonding and stacking. Hydrogen bonding acts transversely to the phosphate backbone bringing bases close to each other. In contrast, stacking acts longitudinally along the backbone, which is the natural direction of the single stranded polymer. It is this geometric alignment between the stacking force and the polymer direction that generates cooperativity. This explains why ssDNA and ssRNA show a cooperative plateau in the FECs \cite{SEOL2007,mcintosh2014} and many RNAs form tertiary structures with coaxial stacking between different helices (making RNA folding a cooperative process \cite{sattin2008,gracia2018}).

To further check the applicability of the model, we carried out pulling experiments to predict the FECs on another ssDNA ($\sim8$ kb, $\sim 50\%$ GC content) obtained by unpeeling an 8kbp dsDNA \cite{EMBO2009,candellitoolbox2013} in a different optical tweezers instrument \cite{SMITH1996}. In App.~\ref{S:Ap.Borja}, we show the predicted FEC from the model using the $\epsilon$, $\gamma$ parameters derived in our study [Fig.~\ref{fig:3}(b)-(c)] without any fitting procedure. The good match with the experimental data further validates our model's predictive power and the derived energy parameters.

In a previous work \cite{bosco2014elastic}, we found a phenomenological formula for the FEC at different salt conditions. This formula included two parameters, a critical force ($f^{\dagger}$) and a force-width parameter ($\delta$), akin to $\epsilon$ and $\gamma$ in the helix-coil model. The phenomenological formula gives comparable fits to the helix-coil model; however, the former has a limited scope at the level of physical interpretation, especially at low forces where their behaviors differ (as shown in Sec. S10 in the Supplemental Material at [URL will be inserted by publisher]).

A remarkable feature of the ssDNA FECs is the absence of force rips, which is due to the low values of $\epsilon$ ($\lesssim$ 0.3kcal/mol) compared to $k_{\rm B}\,T$ ($\sim$0.6kcal/mol at $T=298$K). For a random sequence, Brownian forces smear out any force rips appearing during the formation of compact domains. This gives rise to the smooth and reversible FECs observed in the experiments. For $\epsilon\gtrsim k_{\rm B}\,T$ we should observe FECs with specific sequence-dependent force rips as it is the case in unzipping experiments [Fig.~\ref{fig:1}(b)] where base-pair energies are $\sim 2-4 k_{\rm B}\,T$.  Still, sequence effects are observed in the height of the shoulder of the FEC that increases with GC content. Concomitantly, the value of $\epsilon$ increases too (App.\ref{S:Ap.hairpin_diameter}).

Regarding salt dependence effects, we observe that $\epsilon$ and $\gamma$ have logarithmic salt corrections. The salt-correction parameter for $\epsilon$ in sodium [$m_{\epsilon}=0.11(2)$ kcal/mol] is compatible with that of dsDNA hybridization [$m=0.11(1)$ kcal/mol]. This indicates that monovalent cations play a similar role in secondary structure formation and dsDNA hybridization. The stabilization effect of magnesium appears to be slightly larger for ssDNA folding [$m_{\epsilon}=0.09(2)$ kcal/mol] as compared to duplex hybridization [$m=0.06(2)$ kcal/mol]. This demonstrates the distinct role of divalent cations in bringing together distant nucleotides to fold ssDNA. Such a long-range effect should be less prominent for duplex formation where the stem always grows in the proximity of the hybridization junction. 

Apparently, the interfacial energy, $\gamma$, decreases much weakly with  the ionic strength, $|m_{\gamma}|\ll m_{\epsilon}$ raising the question whether the logarithmic dependence of Eq.~(\ref{eq:gammasalt}) is the right phenomenological description. It would be very interesting to extend thermodynamic activity theories to describe salt corrections to the cooperativity term. Using the less-accurate WLC model to fit the data,  a systematic salt dependence in both sodium and magnesium that is compatible with Eq.~(\ref{eq:gammasalt}) is obtained (see Fig. S6 in the Supplemental Material at [URL will be inserted by publisher] for details). This would be an indication that salt similarly screens the interaction between adjacent bases in secondary structure and dsDNA hybridization.

The model predicts the free energy of the folded structure and its nature at zero force ($\phi_{\textrm{C}}(0)$ and $n_{\textrm{C}}(0)$). Free energies are in agreement with Mfold predictions for all investigated lengths, while $\phi_{\textrm{C}}(0)$ and $n_{\textrm{C}}(0)$ are not well captured for long molecules ($N\gtrsim 1000$b) due to the underestimation of large C-domains ($\sim$ 100b) by our model. Interestingly, the model also predicts that ssDNA folds at zero force even for very short lengths $N\sim10$-$100$b at physiological conditions (10 mM \ce{MgCl2}) [i.e. $\phi_{\textrm{C}}(0)\sim 0.5$ in Figs.~\ref{fig:5}(a) (top), \ref{fig:6}(d)]. This prediction is of biological relevance because short ssDNA fragments ($15-30$b) play important roles during DNA metabolic processes. 

The agreement between the helix-coil model and Mfold is remarkable. This suggests that Mfold includes cooperativity as an essential ingredient for secondary-structure prediction. Indeed cooperativity is implicit in the energy of the stem-loop structures formed by stacking adjacent base pairs in the NN model and other secondary motifs. Mfold does not account for base-pair mismatches and tertiary motifs in general.  This limitation of Mfold excludes the possibility of predicting disordered structures made of a large number of small folded regions of low thermodynamic stability but entropically favorable because of their large number. On the other hand, our model does not include heterogeneity in the sequence while Mfold does, predicting large secondary domains. These two features make the most important differences between the helix-coil model and Mfold.

Thermal molecular folding (at zero force) cannot be directly measured with force spectroscopy. However techniques such as hydrogen exchange combined with NMR and mass spectrometry might grant access to naturally folded domains and folding kinetics \cite{Konermann2011,Largy2020}. The kind of structures expected in ssDNA should differ from those reported in native RNAs and proteins \cite{Woodside2015,Englander2014}. Still, the cooperativity mechanisms might be similar and ultimately related to the fundamental nature of electrostatic interactions \cite{Anthony2012,Zhou2018}.

It might be interesting to apply the current model to investigate ssRNA folding, where base stacking interactions are stronger. Future work should also extend this study to incorporate sequence disorder while keeping the simplicity of a few energy parameters. This might be useful for models of folding in native RNAs and proteins showing cooperativity-dependent sequential folding. 

\section{Acknowledgments}
X.V., M.M. and F.R. acknowledge support from European Union’s Horizon 2020 Grant No. 687089, Spanish Research Council Grants FIS2016-80458-P, PID2019-111148GB-I00. F. R. also acknowledges support from ICREA Academia Prizes 2013 and 2018. M.M. also acknowledges support from the Spanish Ramon y Cajal programme of MICINN. C.R.P and B.I. acknowledge support from Spanish Research Council (PGC2018-099341-B-I00) and Comunidad de Madrid (PEJD-2016/IND-2451).

\appendix
\section{Obtaining ssDNA extension}
\label{S:Ap.ssDNA}

\subsection{Trap position shift determination}
\label{S:Ap.ssDNA_long}
\begin{figure}[h]
\centering
\includegraphics{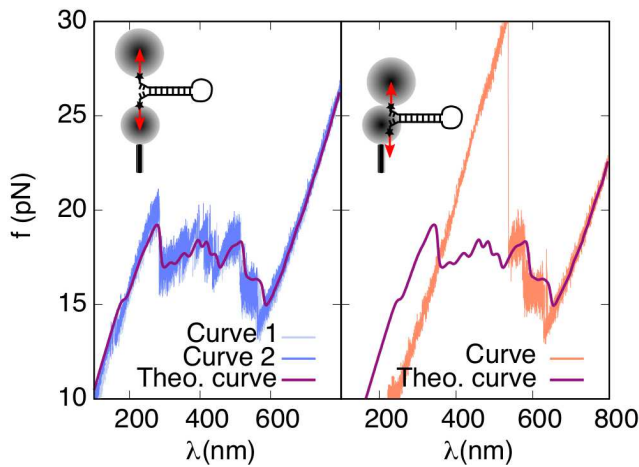}
\caption{{\bf Checking tether alignment.} Experimental force-distance unzipping curves obtained for an aligned tether (left, blue) and a misaligned tether (right, orange). The agreement between the theoretical curve (purple) and the experimental one (blue)is a signature of correct tether alignment  (left panel).
Misaligned tethers usually lead to the lack of some of the initial characteristic unzipping rips (right panel). A schematic depiction of aligned and misaligned tethers are shown on top. For obtaining the theoretical curves, we use the NN energies for the base pair energies \cite{Huguet2010PNAS}, the elastic parameters from \cite{bosco2014elastic} and a bead stiffness of $k_b\sim 0.07$ pN/nm. }
\label{fig:S:theoretical_tether}
\end{figure}

Molecular extension is obtained as described in Eq.~(\ref{eq:lambda_long}), where $x_0$ is a shift of the trap position, $\lambda$. $x_0$ results from imposing $\lambda=0$, at zero force by extrapolating the force-distance curve (FDC) in the initial slope (i.e. before the first rip) to zero force (left scheme in Fig.~\ref{fig:S:theoretical_tether}). For this procedure to work, tethers must be aligned along the pulling axis. To verify the tether alignment, we compare the pattern of the experimental unzipping curve with the theoretically predicted FDC in equilibrium \cite{Huguet2010PNAS}, as shown in Fig~\ref{fig:S:theoretical_tether}. If both curves match, the alignment is correct.

\subsection{Two-branches method}
\label{S:Ap.ssDNA_short}
\begin{figure}[t]
\centering
\includegraphics{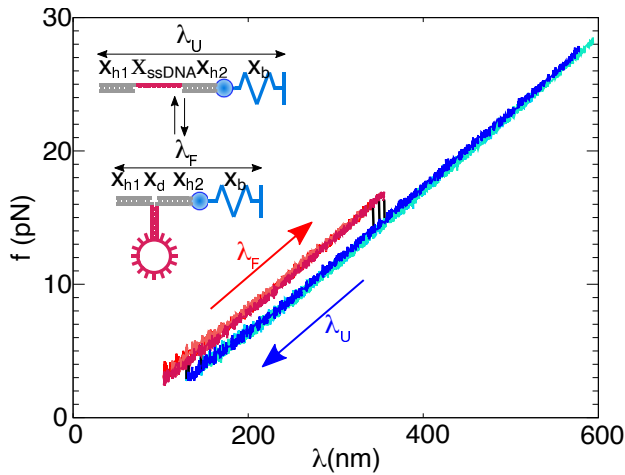}
\caption{{\bf Two-branches method.} Force-distance curves for the $\textrm{H}_{120}$, with several stretching/releasing cycles. Experimental configurations corresponding to F and U branches are schematically shown in the top. The folded ($\lambda_F$) and unfolded ($\lambda_U$) branches are shown in red and blue, respectively, with unfolding and refolding events shown in black. 
}
\label{fig:S:2branches}
\end{figure}

The two-branches method is used for obtaining the ssDNA FEC of the shortest molecule (H$_{120}$). As shown in Fig.~\ref{fig:S:2branches}, the molecule is unzipped and rezipped with pulling velocities of $\sim 100$ nm/s. Two branches are defined depending on whether the hairpin is folded (F) or unfolded (U). The optical trap position in the F ( $\lambda_{F}$) and U ($\lambda_U$) branches, reads as:
\begin{equation}
\lambda_{F}(f)=x_{h}(f)+x_{b}(f)+x_{d}(f)+\lambda_{0,F}.
\label{S:Ap.eq:lambdaF}
\end{equation}
\begin{equation}
\lambda_{U}(f)=x_{h}(f)+x_{b(f)}+x_{\rm ssDNA}(f)+\lambda_{0,U},
\label{S:Ap.eq:lambdaU}
\end{equation}
where $x_d$ is the extension of the folded hairpin and $\lambda_{0,F}$, $\lambda_{0,U}$ are drift corrections.

$x_{\rm ssDNA}(f)$ can be obtained along the force range where the two branches coexist ($\sim$4-17 pN) from Eqs.~\ref{S:Ap.eq:lambdaU} and \ref{S:Ap.eq:lambdaF}:
\begin{equation}
x_{\rm ssDNA}(f)=\lambda_{F}(f)-\lambda_{U}(f)+x_{d}(f)+\lambda_0(t),
\label{S:Ap.eq:ssDNA_short}
\end{equation}
where $\lambda_0(t)$ stands for the drift correction (See Sec. S11 in the Supplemental Material at [URL will be inserted by publisher] for details).

\section{Experiments with glyoxal}
\label{S:Ap.glyoxal}
Experiments with chemically treated ssDNA have been performed in order to check the characterization of the free ssDNA elasticity. Following references \cite{Croquette2001,Saleh2009,McIntosh2011} we have carried out experiments with glyoxal using the following protocol. We first unzipped dsDNA in the absence of glyoxal to generate ssDNA. Since glyoxal prevents base pairing the blocking loop oligonucleotide was not necessary to prevent hairpin rezipping. Once unzipped we kept the force constant above the unzipping force ($\sim$20 pN at 1M NaCl and 10mM MgCl2, $\sim$15pN at 10mM NaCl and 0.5mM MgCl2) and flowed glyoxal at 1M concentration. 

\begin{figure}[h]
\centering
\includegraphics{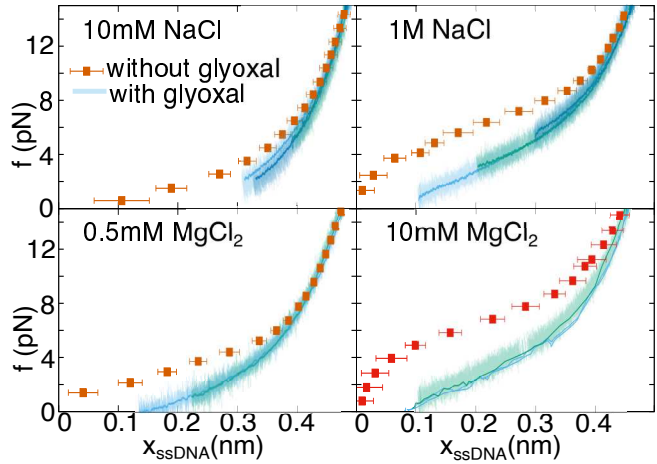}
\caption{{\bf ssDNA FECs with glyoxal.} Re-scaled force-extension curves for different salt concentrations in presence and absence of glyoxal. The red points correspond to the re-scaled averaged FEC for the $\textrm{H}_{13680}$ ssDNA molecule. The blue and green lines show releasing cycles for different tethers of the $\textrm{H}_{7138}$ ssDNA molecule after the secondary structure is suppressed by glyoxal. Experiments were performed in 10mM NaCl (top left), 1M NaCl (top right), 0.5mM \ce{MgCl2} (bottom left) and 10mM \ce{MgCl2} (bottom right). The dark lines show the averaged signal using a 10Hz filter.
\label{fig:Ap.glyoxal}
}
\end{figure}

We waited for 15 minutes for glyoxal to covalently attach to the exposed ssDNA bases. Glyoxal changes the index of refraction of the solvent, modifying force calibration. Therefore, once glyoxal has coated ssDNA we washed the chamber by flowing the original buffer to remove glyoxal and measure the glyoxal-ssDNA FECs. Figure~\ref{fig:Ap.glyoxal} shows the results we obtained for different molecules at the lowest and highest salt conditions in sodium and magnesium. For comparison, we also show the results obtained without glyoxal. The differences between the FECs obtained in the two conditions (with and without glyoxal) is due to the presence or absence of secondary structure.

\begin{figure}[h]
\centering
\includegraphics{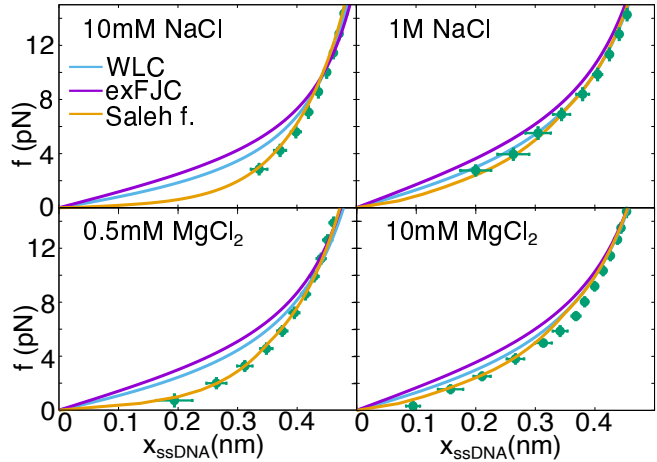}
\caption{{\bf Comparison between elastic models and the experimental FEC with glyoxal.} Green points show the averaged re-scaled FECs of the ${\rm H}_{7138}$ ssDNA molecule for different salt concentrations (from 2-3 different tethers). The continuous lines correspond to the elastic curve obtained from the WLC (blue, elastic parameters from \cite{bosco2014elastic}), extensible Freely Jointed Chain (magenta, elastic parameters from Ref.~\cite{bosco2014elastic}) and orange (Saleh formula, described in Sec.~\ref{subsec:saleh}).
\label{fig:Ap.elasticity_glyoxal}
}
\end{figure}

In Figure~\ref{fig:Ap.elasticity_glyoxal} we show the FECs averaged over the molecules at each condition together with three elastic models: an extensible-FJC model (with parameters from \cite{bosco2014elastic}), the WLC model in Eq.~(\ref{eq:wlc_marko_siggia}) and the Saleh formula in Eq.~(\ref{eq:saleh_formula}). The results show that at high salts the inextensible-WLC and the Saleh formula are both compatible with the data, however there are differences at low salts where the Saleh formula works better. This confirms the results shown in Fig.~S4 in the Supplemental Material at [URL will be inserted by publisher] where the largest differences among the elastic models appear at the lowest salt concentrations.

\section{Hairpin orientation}
\label{S:Ap.hairpin_orient}
A DNA hairpin has a helix diameter of $d=2$ nm, corresponding to that of dsDNA (B-form) \cite{sinden1994dna}. Under an externally applied force, the hairpin extension can be modelled as a single bond of length $d$ that orients along the stretching direction \cite{Woodside2006science,forns2012handles,alemany2014determination}:

\begin{equation}
x_d(f)=d\left( \coth\left( \frac{d f}{k_BT} \right)-\frac{k_BT}{d f}\right),
\label{eq:hairpin_orient}
\end{equation}
where $k_{B}$ is the Boltzmann constant and $T$ is the temperature.

\section{C-domain length}
\label{S:Ap.hairpin_diameter}

\begin{figure}[h]
\centering\includegraphics{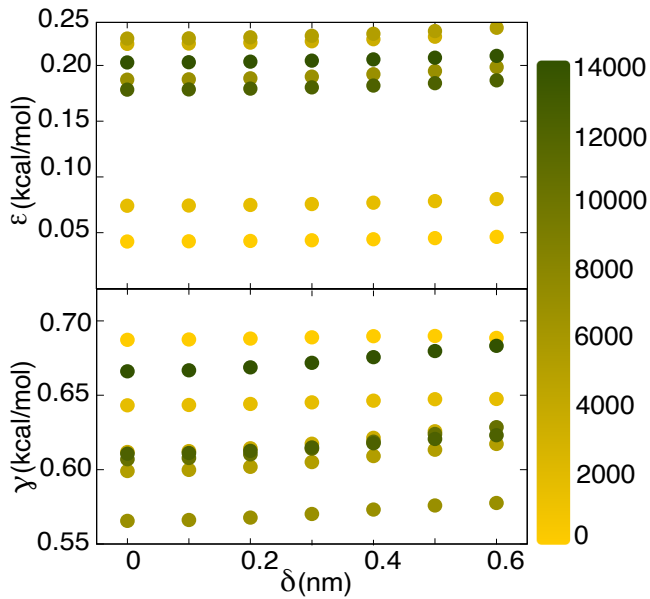}
\caption{{\bf Effect of the domain size in the fitting parameters.} Best fitting values for the model parameters,  $\epsilon$ and  $\gamma$, as a function of the length $\delta$. The effect is very small: $\epsilon$ ($\sim3\%$ variation, top) and $\gamma$ ($\sim3\%$ variation, bottom). The color code indicates the molecular length, from short (yellow) to long (dark green).}
\label{fig:Ap.hairpin_diameter}
\end{figure}

C-domains are modeled as rigid dipoles of length $\delta$ that orient in the presence of force [Eq.~(\ref{eq:hairpin_orient}), $d=\delta$]. By assuming that terminal bases in C-domains form a base pair [as in hairpin-like secondary structures, Fig.~\ref{fig:6}(a)] then $\delta\sim 2$ nm, the double-helix diameter.

We investigated the effect of varying $\delta$ in our model. Starting from $\delta=0$ and upon increasing its value we found that the fitting energy parameters $\epsilon$ and $\gamma$ remain almost unchanged (Fig.~\ref{fig:Ap.hairpin_diameter}): $\epsilon$ and $\gamma$ increase $\lesssim3\%$ when $0\leq \delta \leq 0.6$ nm. We conclude that the effect of $\delta$ is very small in the obtained fitting parameters. For the sake of simplicity, we have used $\delta=0$nm throughout the analysis presented in this work.

Notice that our model, as presented in Sec.\ref{sec:2S-model}, is only applicable for $\delta \leq l\sim 0.7$ nm, otherwise for $\delta>l$ the model predicts, at sufficiently high forces, a structure made of alternating C and F-domains of only one base ($\phi_C=1/2,\,n_C=1$). This is clearly incorrect as secondary structure is fully disrupted at sufficiently high forces. This inaccuracy of the helix-coil model is consequence of its simplicity, and might be corrected by imposing a minimum number of bases, $n_{\rm min}$, to form a C-domain such that $n_{\textrm{C}}>n_{\rm min}$. With this restriction, the helix-coil model becomes too complicated and is not analytically solvable anymore. This fact and the negligible effect of $\delta$ on $\epsilon$, $\gamma$ (Fig.~\ref{fig:Ap.hairpin_diameter}) explain why we took $n_{\rm min}=1$ and $\delta=0$, keeping the simplicity of our model.

\section{GC content effect on secondary structure}
\label{S:Ap.GC}
\begin{figure}[t]
\centering\includegraphics{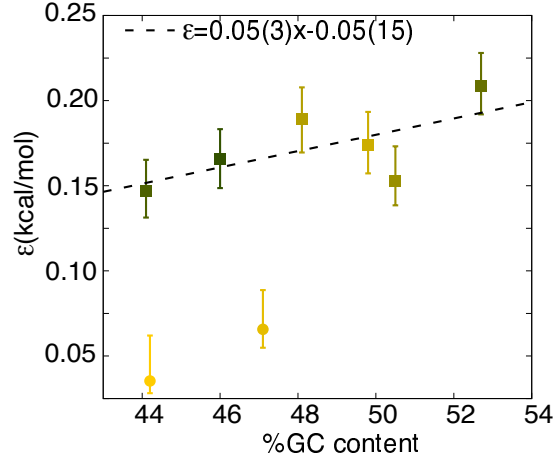}
\caption{{\bf Effect of the GC content in $\epsilon$.} Best fitting values for $\epsilon$ as a function of the GC content of the sequence, for the 8 molecules studied. Long molecules ($N\gtrsim 500$b) are shown in squares, whereas short molecules ($N\lesssim 500$b), deviating from the collapsed FEC trend observed in Fig.~\ref{fig:1}(d), are shown in circles. 
The linear fit (dashed line) performed with long molecules ($700b\leq N \leq 14kb$), shows an increase of $\epsilon$ with GC content, as expected from the higher stability of GC bps as compared to AT bps. Color code as in Fig~\ref{fig:Ap.hairpin_diameter}.}
\label{fig:S:GC_content}
\end{figure}

Sequences with high GC content present the shoulder in the FEC at larger forces \cite{Croquette2001}. The molecules studied in this work (Sec.~\ref{subsec:length_effects}) present small differences in the GC content (44-53\%). For long molecules ($N\gtrsim 500$b, squares), $\epsilon$ as a function of the GC content increases  (Fig.~\ref{fig:S:GC_content}), compatible with the larger stability of secondary structures with high GC content, as shown in Fig.~\ref{fig:S:GC_content}. For shorter molecules ($N\lesssim 500$b, yellow circles), $\epsilon$ decreases strongly, which we interpret as finite size effects (see main text).

\section{Model validation}
\label{S:Ap.Borja}
\begin{figure}
\centering\includegraphics{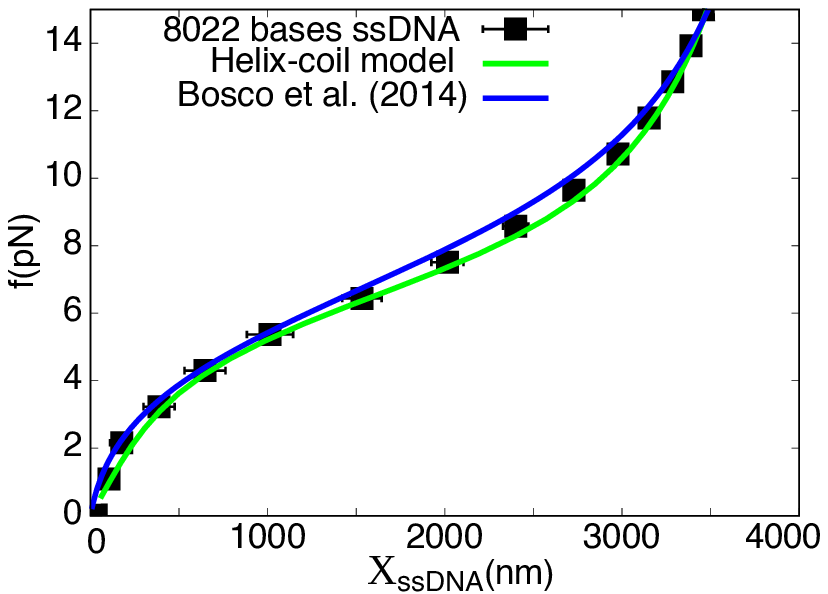}
\caption{{\bf Comparison of the model with a ssDNA FEC obtained using unpeeling of dsDNA.}  Average FEC of a 8022 bases ssDNA (black points) in 20 mM Tris pH7.5, 10 mM \ce{MgCl2} and 20 mM \ce{NaCl}. The errors are the standard errors obtained from averaging three different tethers. The experimental extension of each ssDNA tether is corrected with a small multipying factor (from 0.9 to 1.1) to match the WLC elasticity at 15pN. The FEC predicted by the model (green line), is obtained using Eq.~(\ref{eq:extension_model}) with $\epsilon$ given by Eq.~(\ref{eq:energy_frac_length}) (main text) and $\gamma=0.61$ kcal/mol (average value) [Fig.~\ref{fig:3}(c)]. The blue curve is the phenomenological formula proposed in \cite{bosco2014elastic} (See Sec.~S10 in the Supplemental Material at [URL will be inserted by publisher] for the used parameters).}
\label{fig:S:Borja}
\end{figure}

To test the model, we used experimental FEC data of a 8022 bases ssDNA molecule obtained by unpeeling a dsDNA molecule using a different optical tweezers setup~\cite{SMITH1996} (Fig.~\ref{fig:S:Borja}). In this method, one strand of the dsDNA molecule is attached by its 5' and 3'ends to the beads. Pulling the molecule above 80-100 pN ($\sim200$ nm/s) induces the mechanical denaturalization of the double helix and promotes the releasing of the free strand (the one not attached to the beads) \cite{EMBO2009,candellitoolbox2013}.
We use our model without any fitting parameters to predict the FEC of this molecule at the experimental conditions (10 mM \ce{MgCl2} and 20 mM \ce{NaCl}). The good match between the model prediction and the experimental data (Fig.~\ref{fig:S:Borja}) further validates the model.

\section{Helix-coil model derivation}
\label{S:Ap.2SModel}

The model Hamiltonian, given by Eq.~(\ref{eq:hamiltonian_1}), can be written as:

\begin{equation}
 \mathcal{H}=-A\sum_i^{N}\sigma_i-B\sum_i^{N}\sigma_i\sigma_{i+1}-C(f)-D,
 \label{eq:hamiltonian_general}
\end{equation}
where

\begin{eqnarray}
A&=&  -\frac{\epsilon}{2}+\frac{1}{2}\int_0^{f}x_{\textrm{F}}(f')df', \label{eq:termsA}\\
B&=& -\frac{1}{4}\int_0^{f}d_{\textrm{C}}(f')df'+\gamma,  \label{eq:termsB}\\
C&=& \frac{N}{2}\left( \int_0^{f}x_{\textrm{F}}(f')df'+\frac{1}{2}\int_0^{f}d_{\textrm{C}}(f')df'\right), \label{eq:termsC} \\
D&= &\frac{N}{2}\epsilon.
\label{eq:termsD}
\end{eqnarray}
$D$ is a constant, and therefore will not be considered in the calculations below.
The partition function reads as: 

\begin{equation}
Z=\sum_{\{\sigma_i\}}e^{-\beta \mathcal{H}(\{\sigma_i\})}=\sum_{\sigma_1}V^N\sigma_1\sigma_1=tr\left(V^N\right),
    \label{eq:partition_function_matrix}
\end{equation}
where $tr$ is the trace and $V$ is the transfer matrix given by:
\begin{equation}
 V=e^{C/N}
 \begin{pmatrix}
    e^{\beta(A+B)} & e^{-\beta B} \\
    e^{-\beta B} & e^{-\beta(A-B)} 
  \end{pmatrix}.
 \label{eq:transfer_matrix1}
\end{equation}
By diagonalizing $V$, we find the eigenvalues:
\begin{eqnarray}
 \lambda_{\pm}&=&e^{\beta C/N}e^{\beta B} \cdot \nonumber \\
 &&\cdot \left[ \cosh\left(\beta A\right)\pm\sqrt{e^{-4\beta B}+\sinh^2\left(\beta A\right)}\right]. \\
 \label{eigenvalues}
\end{eqnarray}
The partition function [Eq.~(\ref{eq:partition_function_matrix})] can then be written as:
\begin{equation}
Z_{N}=\lambda_{+}^N\left[1+\left(\frac{\lambda_{-}}{\lambda_{+}}\right)^N \right],
 \label{eq:partition_function}
\end{equation}
which, in the thermodynamic limit $N\to\infty$ ($\lambda_{-}<\lambda_{+}$), leads to: 
\begin{equation}
Z_{N}=\lambda_{+}^N.
 \label{eq:partition_function_simple}
\end{equation}

\subsection{Free-energy computation}

From the partition function [Eq.~(\ref{eq:partition_function_simple})],
we can compute the free-energy per monomer (base) as: 

\begin{eqnarray}
&\frac{G}{N}&=-\frac{1}{\beta}\log\left(\lambda_{+}\right)= -\frac{C}{N}-B- \nonumber\\
&-&\frac{1}{\beta}\log\left[\cosh\left(\beta A\right)+\sqrt{e^{-4\beta B}+\sinh^2\left(\beta A\right)}\right],
 \label{eq:free-energy1}
\end{eqnarray}
with  $A$, $B$ and $C$ given in  Eqs.~(\ref{eq:termsA}-\ref{eq:termsC}). The free-energy per monomer at zero force, $\frac{G_0}{N}$, is obtained by imposing $f=0$ in Eq. ~(\ref{eq:free-energy1}):
\begin{eqnarray}
&\frac{G_0}{N} &=-\gamma-\frac{1}{\beta}\log\Bigl[\cosh\left(-\beta \epsilon/2\right)+\nonumber \\
&+&\sqrt{e^{-4\beta \gamma}+\sinh^2\left(-\beta \epsilon/2\right)}\Bigr].
 \label{eq:free-energy_zero_force}
\end{eqnarray}

For a given DNA sequence, Mfold gives the free energy of formation of secondary structures with respect the random coil (i.e. configuration with all bases free).  In order to compare the free-energy of the helix-coil model with Mfold predictions, we subtract to Eq.~(\ref{eq:free-energy_zero_force}) the energy per base of the random coil at zero force  [Eq.~(\ref{eq:hamiltonian_1}) with $f=0$, $\sigma_i=1, \forall i$], $\epsilon/2-\gamma$:

\begin{eqnarray}
 \frac{\Delta G_0}{N}=-\frac{\epsilon}{2}-\frac{1}{\beta}\log\Bigl[\cosh\left(\frac{\beta\epsilon}{2}\right)+ \nonumber\\
 \sqrt{e^{-4\beta\gamma}+\sinh^{2}\left(\frac{\beta\epsilon}{2}\right)}\Bigr].
\end{eqnarray}

\subsection{Computation of $\phi_{F}(f)$, $\phi_C(f)$ and $M_C(f)$}
\label{sec:calculation_n1}
The fraction of bases in F-domains can be written as:
\begin{equation}
 \phi_{\textrm{F}}(f)=\frac{N_F(f)}{N}=\frac{1+\left \langle \sigma_n \right \rangle(f)}{2},
 \label{eq:bases1}
\end{equation}
where $\left \langle \right \rangle$ stands for the ensemble average.

Using the transfer matrix definition in Eq.~(\ref{eq:transfer_matrix1}), the average $\sigma$ for the first monomer reads as:
\begin{eqnarray}
\left \langle \sigma_1 \right \rangle(f)&=\frac{1}{Z}\sum_{\sigma_1...\sigma_N}\sigma_1V(\sigma_1,\sigma_2)V(\sigma_2,\sigma_3)...V(\sigma_N,\sigma_1)=\nonumber\\
&=\frac{1}{Z}\sum_{\sigma_1...\sigma_N}V'V^{N-1},
\label{eq:magnetization1}
\end{eqnarray}
where 
\begin{equation}
 V'=e^{C/N}
 \begin{pmatrix}
    e^{\beta(A+B)} & e^{-\beta B} \\
    -e^{\beta B} & -e^{-\beta(A-B)} 
  \end{pmatrix},
 \label{eq:transfer_matrix2}
\end{equation}
and  $A$, $B$ and $C$ are given in  Eqs.~(\ref{eq:termsA}-\ref{eq:termsC}). Considering Pauli matrices, $V'=\sigma_zV$, we can write:
\begin{equation}
\left \langle \sigma_1 \right \rangle(f)=\frac{1}{Z}tr\left(\sigma_z V^N\right).
\label{eq:magnetization2}
\end{equation}
Using the cyclic property of the trace, Eq.~(\ref{eq:magnetization2}) can be extended to any site, $n$: 
\begin{equation}
\left \langle \sigma_n \right \rangle(f)=\frac{1}{Z}tr\left(\sigma_z V^N\right).
\label{eq:magnetization3}
\end{equation}

By diagonalizing $V$, we can write: 
\begin{equation}
\left \langle \sigma_n \right \rangle(f)=\frac{1}{Z}\left(\langle v_{\pm}\lvert\sigma_z\lvert v_{\pm}\rangle\lambda_+^N+\langle v_{\pm}\lvert\sigma_z\lvert v_{\pm}\rangle \lambda_-^N \right),
\label{eq:magnetization4}
\end{equation}

where $\lvert v_{\pm}\rangle$ are the eigenvectors: 

\begin{eqnarray}
\lvert v_+\rangle&=
  \begin{pmatrix}
    \cos\left(\theta/2\right) \\
    \sin\left(\theta/2\right)
  \end{pmatrix}, \nonumber\\
\lvert v_-\rangle&=
  \begin{pmatrix}
    -\sin\left(\theta/2\right) \\
    \cos\left(\theta/2\right) 
  \end{pmatrix},
 \label{eq:eigenvector1}
\end{eqnarray}

with the angle $\theta$ given by:
\begin{equation}
 \tan\theta=\frac{e^{-2\beta B}}{\sinh\left(\beta A\right)}.
 \label{eq:angle}
\end{equation}
In the thermodynamic limit $Z=\lambda_+^N$ and Eq.~(\ref{eq:magnetization4}) reduces to: 

\begin{equation}
\left \langle \sigma_n \right \rangle(f)=\langle v_{\pm}\lvert\sigma_z\lvert v_{\pm}\rangle.
\label{eq:magnetization5}
\end{equation}

By substituting Eq.~(\ref{eq:eigenvector1}) to Eq.~(\ref{eq:magnetization5}), we obtain:
\begin{eqnarray}
\left \langle \sigma_n \right \rangle(f)&=\cos^2\left(\theta/2\right)-\sin^2\left(\theta/2\right)=\cos\left(\theta\right)=\nonumber \\
&=\frac{\sinh\left(\beta A\right)}{\sqrt{e^{-4\beta B}+\sinh^2\left(\beta A\right)}}.
\label{eq:magnetization6}
\end{eqnarray}
Finally, using Eq.~(\ref{eq:magnetization6}) in Eq.~(\ref{eq:bases1}) leads to:

\begin{equation}
 \phi_{\textrm{F}}(f)=\left(\frac{1}{2}+\frac{\sinh\left(\beta A\right)}{2\sqrt{e^{-4\beta B}+\sinh^2\left(\beta A\right)}}\right).
 \label{eq:prob1}
\end{equation}

The fraction of bases in C-domains $\phi_C(f)=\left(1-N_{\textrm{F}}(f)\right)/N$, then reads as: 

\begin{equation}
 \phi_{\textrm{C}}(f)=\left(\frac{1}{2}-\frac{\sinh\left(\beta A\right)}{2\sqrt{e^{-4\beta B}+\sinh^2\left(\beta A\right)}}\right).
 \label{eq:prob-1}
\end{equation}

On the other hand, the number of compact domains, $M_{\rm C}$, is given by: 
  \begin{equation}
 \langle M_{\rm C} \rangle =\frac{N}{4}\left(1-\langle \sigma_i\sigma_{i+1}\rangle\right).
 \label{eq:number_hairpins_3}
\end{equation}

Using the transfer matrix formalism, we can write $\langle \sigma_n\sigma_{n+r}\rangle$ as:

\begin{eqnarray}
\left \langle \sigma_n\sigma_{n+r} \right \rangle(f)&=\frac{\langle v_+\lvert \sigma_z V^r \sigma_z\lvert v_+\rangle}{\langle v_+\lvert  V^r \lvert v_+\rangle}=\nonumber \\
&= \cos^2\left(\theta\right)+\left(\frac{\lambda_-}{\lambda_+}\right)^r\sin^2\left(\theta\right).
\label{eq:correlation1}
\end{eqnarray}

The number of compact domains [Eq.~(\ref{eq:number_hairpins_3})], then reads as:
\begin{eqnarray}
 \langle M_{\rm C} \rangle &=\frac{N}{4}\sin^2\theta\left(1-\frac{\lambda_-}{\lambda_+}\right) =\nonumber \\
&=\frac{N}{2}\frac{\sinh\left(\beta A\right)}{\cosh\left(\beta A\right)+\sqrt{e^{-4\beta B}+\sinh^2\left(\beta A\right)}},
 \label{eq:number_hairpins_4}
\end{eqnarray}
where we have used Eq.~(\ref{eq:magnetization6}).

\section{DNA hairpin synthesis}
\label{S:Ap.DNAsynth}

\begin{figure}[h!]
\centering
\includegraphics[width=\linewidth]{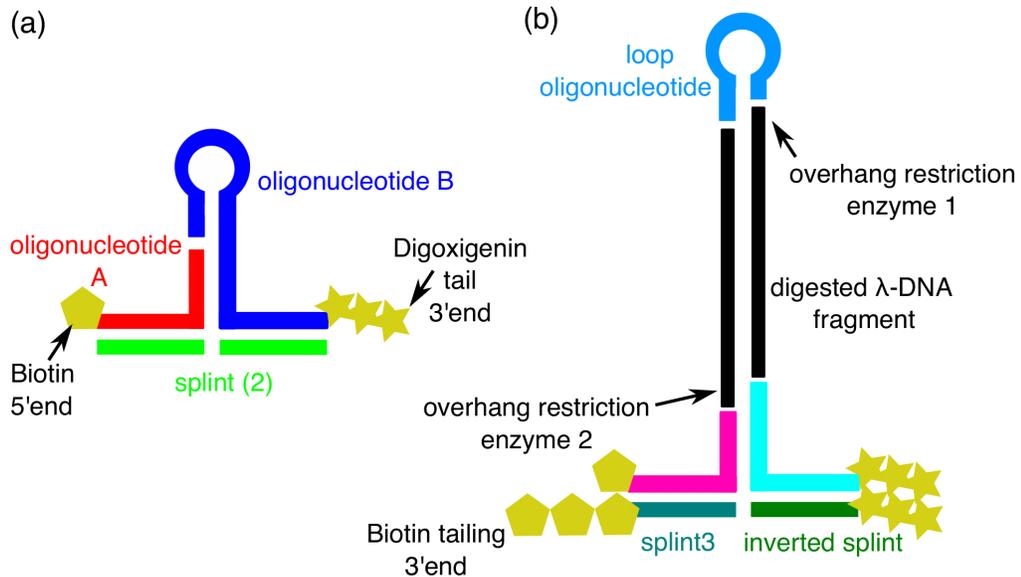}
\caption{{\bf Hairpin synthesis.} (a) $\textrm{H}_{120}$ is assembled by ligating two oligonucleotides (blue and red). The oligonucleotide A is purchased biotinylated and the oligonucletide B is end-labelled with digoxigenins using the T4 terminal transferase. The splint oligonucletides are annealed to create dsDNA handles. $\textrm{H}_{204}$ is synthesized following the same steps, but the initial assembling involves 4 oligonucleotides (A, B, C, D). 
(b) The long DNA hairpins are assembled by ligating a set of oligonucleotides (magenta, cyan, blue) to the PCR-amplified and digested $\lambda$-phage fragment (black). Note that the complementary strands of the handles are also tailed. Color code as in Tables~\ref{Ap.table:oligos_short} and \ref{Ap.table:oligos_long}.  }
\label{fig:S:hairpin}
\end{figure}

$\textrm{H}_{120}$ and $\textrm{H}_{204}$ were synthesized following the protocol in Ref.~\cite{alemany2014determination}. The oligonucleotide forming the 3' end of the hairpin was labelled with a digoxigenin tailing. After a purification step (\textit{QIA Nucleotide removal kit}), all the oligonucleotides forming the hairpin (Table \ref{Ap.table:oligos_short}) were annealed by starting at a high temperature (\SI{70}{\celsius}) and \SI{1}{\celsius} was decreased every minute until room temperature was reached. The hairpin was next ligated using the T4 DNA ligase (New England Biolabs) in an overnight reaction (\SI{16}{\celsius}). 
$\textrm{H}_{700}$, $\textrm{H}_{964}$, $\textrm{H}_{4452}$ and $\textrm{H}_{13680}$ were prepared as described in Refs.~\cite{camunas2015footprinting} and \cite{camunas2013electrostatic}.
Finally, $\textrm{H}_{1904}$ and $\textrm{H}_{7138}$ were synthesized following the protocol in Ref.~\cite{camunas2013electrostatic}, changing the restriction enzyme for the digestion step: {\sl EcoRI} (New England Biolabs) for $\textrm{H}_{7138}$, and {\sl BspHI} (New England Biolabs) for $\textrm{H}_{1904}$. The sequences of the oligonucleotides used for preparing the DNA hairpins are given in Sec.~\ref{S:Ap.DNAoligos}. Fig.~\ref{fig:S:hairpin}(a)-(b) shows an scheme of the molecular construct for short ($N\leq204$ b) and long ($N>204$ b) hairpins, respectively.

\clearpage
\newpage

\section{Oligonucleotides for hairpin synthesis}
\label{S:Ap.DNAoligos}
\begin{table}[h!]\footnotesize
\centering
\begin{tabularx}{\textwidth}{s b}
\hline
\textbf{Name} & \textbf{Sequence} \\
\hline
\textcolor{red}{120b-A} & \textcolor{red}{5$'$-Biotin-AGT TAG TGG TGG AAA CAC AGT GCC AGC GCG AAC CCA CAA ACC GTG ATG GCT GTC CTT GGA GTC ATA CGC AA -3$'$}  \\
\textcolor{blue}{120b-B} & \textcolor{blue}{5$'$-GAA GGA TG\textbf{G AAA AAA AAA AAA AAA AAA A}CA TCC TTC TTG CGT ATG ACT CCA AGG ACA GCC ATC ACG GTT TGT GGG TTC AGT TAG TGG TGG AAA CAC AGT GCC AGC GC-3$'$} \\
\textcolor{red}{204b-A} & \textcolor{red}{5$'$-Bio-AGT TAG TGG TGG AAA CAC AGT GCC AGC GCC GAC CTC T-3$'$} \\
\textcolor{blue}{204b-B} & \textcolor{blue}{5$'$-Pho-TAA CCT CCA AGC GTA CAG GGT GGA CTT TGC AAC GAC TTC CTA GAC CAA AGA CTC GCT GTT TAC GAA ATT TGC GCT CAA GCG AGA GTA TTG AA\textbf{T TTT} TTC AAT AC-3$'$}\\
\textcolor{blue}{204b-C} & \textcolor{blue}{5$'$-Pho-TCT CGC TTG AGC GCA AAT TTC GTA AAC AGC GAG TCT TTG GTC TAG GAA GTC GTT GCA AAG TCC ACC CTG TAC GCT TGG AGG -3$'$}\\
\textcolor{blue}{204b-D} & \textcolor{blue}{5$'$-Pho-TTA AGA GGT CGA GTT AGT GGT GGA AAC ACA GTG CCA GCG C -3$'$}\\
\textcolor{green}{splint} & \textcolor{green}{5$'$-GCG CTG GCA CTG TGT TTC CAC CAC TAA CT-3$'$}\\
\hline

\end{tabularx}
\caption{\label{Ap.table:oligos_short}Oligonucleotides used for the synthesis of $\textrm{H}_{120}$ and $\textrm{H}_{204}$. The loop region is shown in bold.}
\end{table}

\begin{table}[h!]\footnotesize
\centering
\begin{tabularx}{\textwidth}{s b}
\hline
\textbf{Name} & \textbf{Sequence} \\
\hline
\textcolor{myblue}{13680b-loop} & \textcolor{myblue}{5$'$-Pho-GAT CGC CAG TTC GCG TTC GCC AGC ATC CG{\bf A CTA} CGG ATG CTG GCG AAC GCG AAC TGG C-3$'$} \\
\textcolor{myblue}{7138b-loop} & \textcolor{myblue}{5$'$-Pho-AAT TGC CAG TTC GCG TTC GCC AGC ATC CG{\bf A CTA} CGG ATG CTG GCG AAC GCG AAC TGG C-3$'$} \\
\textcolor{myblue}{4452b-loop} & \textcolor{myblue}{5$'$-Pho-TGA TAG CCT {\bf ACT A}AG GCT ATC ACA TG-3$'$} \\
\textcolor{myblue}{1904b-loop} & \textcolor{myblue}{5$'$-Pho-CAT GAC AGT CGT TAG TAA CTA ACA TGA TAG TTA C{\bf TT TT}G TAA CTA TCA TGT TAG TTA CTA ACG ACT GT-3$'$}  \\
\textcolor{myblue}{964b-loop} & \textcolor{myblue}{5$'$-Pho-GTC ACT TAG TAA CTA ACA TGA TAG TTA C{\bf TT TT}G TAA CTA TCA TGT TAG TTA CTA A-3$'$} \\
\textcolor{myblue}{700b-loop} & \textcolor{myblue}{5$'$-Pho-GTC ACT TAG TAA CTA ACA TGA TAG TTA C{\bf TT TT}G TAA CTA TCA TGT TAG TTA CTA A-3$'$} \\
\textcolor{mypink}{Bio-cosRshort} & \textcolor{mypink}{5$'$-Bio-GAC TTC ACT AAT ACG ACT CAC TAT AGG GAA ATA GAG ACA CAT ATA TAA TAG ATC TT-3$'$} \\
\textcolor{mycyan}{cosRlong} & \textcolor{mycyan}{5$'$-Pho-GGG CGG CGA CCT AAG ATC TAT TAT ATA TGT GTC TCT ATT AGT TAG TGG TGG AAA CAC AGT GCC AGC GC-3$'$} \\
\textcolor{mypink}{Bio-cosLshort}  & \textcolor{mypink}{5$'$-Bio-GAC TTC ACT AAT ACG ACT CAC TAT AGG GAA ATA GAG ACA CAT ATA TAA TAG ATC TT-3$'$} \\
\textcolor{mycyan}{cosLlong}  & \textcolor{mycyan}{5$'$-Pho-AGG TCG CCG CCC AAG ATC TAT TAT ATA TGA GTC TCT ATT AGT TAG TGG TGG AAA CAC AGT GCC AGC GC 3$'$}\\
\textcolor{mypink}{HandBio-SMFP} &\textcolor{mypink}{ 5$'$-Bio-GAC TTC ACT AAT ACG ACT CAC TAT AGG GAA ATA GAG ACA CAT ATA TAA TAG ATC TTC GCA CTG AC -3$'$}\\
\textcolor{mycyan}{HandDig-SMFP} & \textcolor{mycyan}{5$'$-Pho-AAG ATC TAT TAT ATA TGT GTC TCT ATT AGT TAG TGG TGG AAA CAC AGT GCC AGC GC -3$'$}\\
\textcolor{mydarkgreen}{splint3}  & \textcolor{mydarkgreen}{5$'$-TCC CTA TAG TGA GTC GTA TTA GTG AAG TC-3$'$} \\
\textcolor{mygreen}{inverted-splint} &  \textcolor{mygreen}{3$'$-AAA AA-5$'$-5$'$-GCG CTG GCA CTG TGT TTC CAC CAC TAA C(SpC3)-3$'$}\\
\hline
\end{tabularx}
\caption{\label{Ap.table:oligos_long}Oligonucleotides used for the synthesis of long DNA hairpins. The loop region is shown in bold.}
\end{table}

\begin{table}[h!]
\centering
\begin{tabularx}{\textwidth}{s b}
\hline
\textbf{Name} & \textbf{Sequence} \\
\hline
13680b-block-loop & 5$'$-TAG TCG GAT GCT GGC GAA CGC GAA CTG GCG-3$'$  \\
7138b -block-loop & 5$'$-TAG TCG GAT GCT GGC GAA CGC GAA CTG GCG-3$'$\\
4452b-block-loop & 5$'$-TAG TAG GCT ATC ACA TGC TGG CCA CCG GCT-3$'$ \\
1904b-block-loop & 5$'$-TTA CAA AAG TAA CTA TCA TGT TAG T-3$'$ \\
964b-block-loop & 5$'$-TTA CAA AAG TAA CTA TCA TGT TAG T-3$'$ \\
700b-block-loop & 5$'$-TTA CAA AAG TAA CTA TCA TGT TAG T-3$'$ \\
204b-block-loop & 5$'$-AAA ATT CAA TAC TCT CGC TTG AGC G-3$'$ \\
\hline
\end{tabularx}
\caption{\label{S:Ap.table:oligos_block}Blocking loop oligonucleotides used to generate ssDNA FECs for the  different hairpins.}
\end{table}

\clearpage
\newpage

\section{Drift correction}
\label{S:Ap.Drift}

\begin{figure}[h!]
\centering
\includegraphics[width=\linewidth]{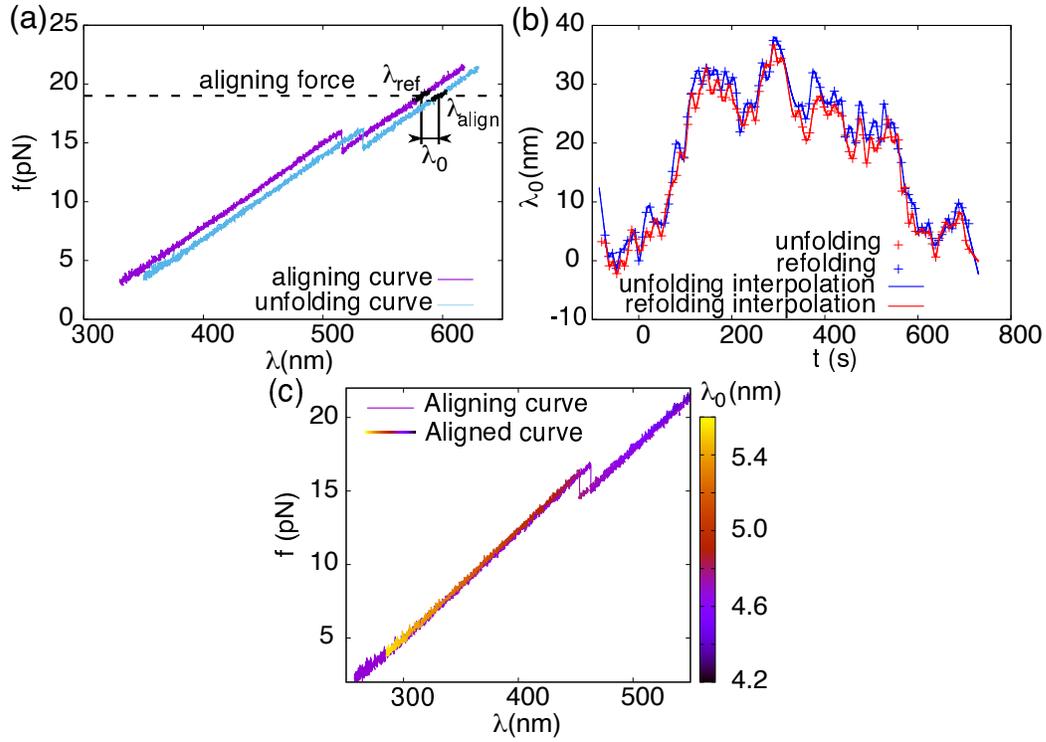}
\caption{{\bf Drift correction.} (a) Two unfolding curves $f(\lambda)$ for $\textrm{H}_{120}$. The magenta curve is the reference curve with $\lambda_{ref}$ at the aligning force $\sim20$ pN, whereas the blue aligning curve has $\lambda_{align}$ at the same force. The shift $\lambda_0=\lambda_{align}-\lambda_{ref}$ is used to correct the molecular extension.     
(b) $\lambda_0$ as a function of time for the unfolding (blue) and refolding (red) curves (crosses). The spline interpolations are shown as solid lines. (c) The two unfolding curves shown in panel (a) after the drift correction has been performed. The aligned curve is plotted using the colour code in the right of the figure for the shift correction $\lambda_0$ applied at each point.}
\label{fig:S:drift}
\end{figure}

The drift is a low frequency noise due to macroscopic effects, such as temperature changes or mechanical vibrations. Correcting the drift is important to precisely determine the molecular extension. For long molecules ($\sim 1000$b), a constant shift $\lambda_0$ is imposed in each folding-unfolding cycle to ensure the collapse between consecutive cycles, as described in \cite{bosco2014elastic}. 
However, for shorter molecules (e.g. $\textrm{H}_{120}$), a time dependent shift, $\lambda_0(t)$, needs to be considered to obtain an accurate estimation of the molecular extension. To do so, we first take a reference unfolding curve [e.g. magenta curve in Fig.~\ref{fig:S:drift}(a)] and find the position of the trap, $\lambda_{ref}$, at a certain aligning force where the hairpin is always unfolded ($\sim20$ pN, black points in the magenta curve). Next, for all cycles of the tether, the position of the trap at the aligning force, $\lambda_{align}$, is computed, and the shift calculated as $\lambda_{0}=\lambda_{align}-\lambda_{ref}$ [Fig.~\ref{fig:S:drift}(a)]. The continuous function $\lambda_{0}(t)$ is build by performing two spline interpolations (for unfolding and refolding curves) to $\lambda_{0}$ as a function of time data, [Fig.~\ref{fig:S:drift}(b)], as in Ref.~\cite{Huguet2010PNAS}. This allows us to perform sub-nanometric corrections, as shown in Fig.\ref{fig:S:drift}(c).

\clearpage
\newpage

\section{Elastic parameters and scaling factors}
\label{S:Ap.scaling}

At high forces (around $\sim15$ pN or above) ssDNA is expected to behave as an ideal polymer. The ideal regime can be described by the WLC model [main text, Eq.~(2)]. Since the scope of the work is not the characterization of this regime, we use the WLC elastic parameters from Ref.~\cite{bosco2014elastic} (Sec.~E in main text) for the bases forming F- domains (i.e. ideal regime) in the different salt conditions. In this way, we can reduce the number of free parameters of the model to just two: $\epsilon$ and $\gamma$. However, this restriction leads to small differences between the experimental FECs and the theoretical WLC curves at high forces. To correct these differences, we impose that the ssDNA extension at $f=15$ pN coincides with that of the WLC model by multiplying the measured extension by a scaling factor, comprised between 0.95 and 1.05. The elastic parameters (from Ref.~\cite{bosco2014elastic}) and scaling factors used for each molecule and salt condition are given in Tables~\ref{Ap.table:scaling_length}, \ref{Ap.table:scaling_nacl} and \ref{Ap.table:scaling_mgcl}.

\begin{table}[h]
\caption{\label{Ap.table:scaling_length} WLC elastic parameters and scaling factors used for different molecules at 10mM ce{MgCl2}.}
\begin{tabular}{l c c c}
Molecule & $p$(nm) & $l$(nm) & Scaling factor \\
\hline
H$_{120}$ & 0.75 & 0.69 & 0.99 \\
H$_{204}$ & 0.75 & 0.69 & 0.95 \\
H$_{700}$ & 0.75 & 0.69 & 0.95 \\
H$_{964}$ & 0.75 & 0.69 & 1.04 \\
H$_{1904}$ & 0.75 & 0.69 & 1.01 \\
H$_{4452}$ & 0.75 & 0.69 & 1.01 \\
H$_{7138}$ & 0.75 & 0.69 & 1.02 \\
H$_{13680}$ & 0.75 & 0.69 & 1.02 \\
\hline
\end{tabular}
\end{table}

\begin{table}[h]
\caption{\label{Ap.table:scaling_nacl} WLC elastic parameters and scaling factors used for H$_{13680}$ for different \ce{NaCl} conditions.}

\begin{tabular}{l c c c}
[\ce{NaCl}](mM) & $p$(nm) & $l$(nm) & Scaling factor \\
\hline
10 & 1.37 & 0.69 & 1.00 \\
25 & 1.12 & 0.69 & 0.97 \\
50 & 0.99 & 0.69 & 0.98 \\
100 & 0.90 & 0.69 & 0.96 \\
250 & 0.82 & 0.69 & 0.96 \\
500 & 0.78 & 0.69 & 0.97 \\
1000 & 0.75 & 0.69 & 0.98 \\
\hline
\end{tabular}
\end{table}

\begin{table}[h]
\caption{\label{Ap.table:scaling_mgcl} WLC elastic parameters and scaling factors used for H$_{13680}$ for different \ce{MgCl2} conditions.}
\begin{tabular}{l c c c}
[\ce{MgCL2}](mM) & $p$(nm) & $l$(nm) & Scaling factor \\
\hline
0.5 & 0.92 & 0.69 & 1.05 \\
1 & 0.86 & 0.69 & 1.04 \\
2 & 0.81 & 0.69 & 1.00 \\
4 & 0.78 & 0.69 & 0.99 \\
10 & 0.75 & 0.69 & 0.96 \\
\hline
\end{tabular}
\end{table}

\clearpage
\newpage

\section{Blocking oligonucleotide method for $\textrm{H}_{204}$ }
\label{S:Ap.oligo_effects}
\begin{figure}[h]
\centering\includegraphics[width=0.7\linewidth]{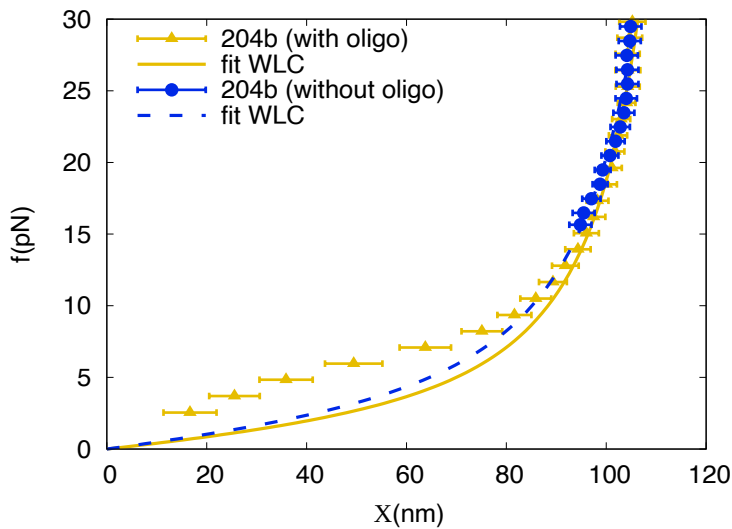}
\caption{{\bf Binding of the blocking oligonucleotide for $\textrm{H}_{204}$.} Average FEC for $\textrm{H}_{204}$, with (triangles, yellow) and without (circles, brown) blocking loop oligonucleotide (25 bases). The values for the best fitting elastic parameters $l$ and $p$ (Eq.~(2), main text) for the FEC with (black, continuous line) and without (brown, dashed line) blocking oligonucleotide are shown in the key.}
\label{fig:S:blocking_oligo_comparison}
\end{figure}

The binding of the blocking oligonucleotide in the loop region slightly changes the elastic response of the molecule. However, the change in the extension per base induced by the hybridized oligonucleotide decreases with the length of the molecule and it is negligible for long enough molecules \cite{bosco2014elastic}.
In order to investigate the effect of the blocking oligonucleotide, we measured the FEC (following the procedure described in App.~A, main text) with and without the blocking oligonucleotide for the shortest molecule studied with this method, $\textrm{H}_{204}$.
As it is shown in Fig.~\ref{fig:S:blocking_oligo_comparison}, both FECs as well as the elastic fits (persistence, $p$, and contour length, $l$) are compatible within error bars. This indicates that the dsDNA hybridized region has a minimal effect on the elasticity of the molecule, for molecules with $N\gtrsim204$.

\clearpage
\newpage

\section{Obtaining trap stiffness}

For forces of the order of a few tens of pN (i.e. small bead displacements in the trap), the optical trap can be approximated as an harmonic potential, with a trap stiffness $k_b$. This allows us to relate the displacement of the bead in the trap, $x_b$, with the applied force as $f = k_b\,x_b$. The bead in the optical trap can be modeled as a Brownian particle in a harmonic potential, and the power spectrum $S_f(\nu)$ of the measured force follows a Lorenzian distribution:

\begin{equation}
    S_f(\nu)=\frac{k_{\textrm{B}}T}{2\pi^2\gamma}\frac{k_b^2}{\nu^2+\nu_c^2},
    \label{eq:lorenzian}
\end{equation}

where $\gamma$ is the drag coefficient of the particle, $\nu$ is the frequency and $\nu_c$ is the so-called corner frequency: $\nu_c=k_b/(2\pi\gamma$). From fitting Eq.~(\ref{eq:lorenzian}) to the zero-force power spectrum, $\gamma$ and $k_b$ are determined.

\begin{figure}[h!]
\centering
\includegraphics{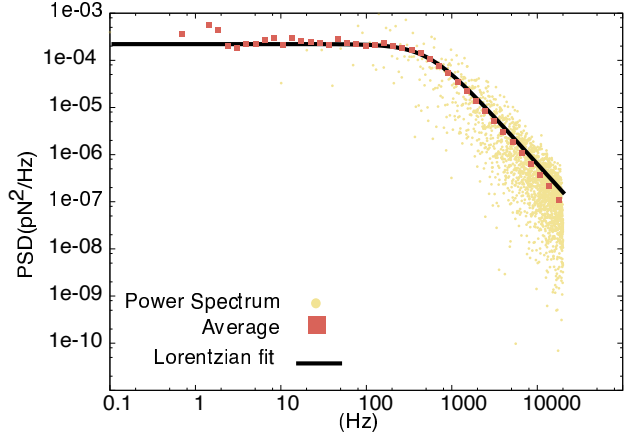}
\label{fig:power_spec}
\caption{{\bf Force power spectrum of a trapped bead at zero force.} Power spectrum obtained from the Fourier transform of the registered force signal (orange) and its log-spaced average (red points) in a log-log representation. The black line shows the fit of Eq.~(\ref{eq:lorenzian}) to the red points, leading to $k_b=0.088\pm0.004$ pN/nm.}
\end{figure}

\clearpage
\newpage

\section{Comparison with the phenomenological formula from \cite{bosco2014elastic}}

In our helix-coil model, the extension of C-domains in the stretching direction can be neglected (App.~C, main text), and therefore the ssDNA extension per monomer Eq.~(12) (main text) can be written in terms of the fraction of bases in F-domains and its extension:

\begin{equation}
 x_{\rm ssDNA}(f)=\phi_{\textrm{F}}(f)x_{\textrm{F}}(f),
  \label{eq:extension_model2}
 \end{equation}
where $x_{\textrm{F}}(f)$ is the extension of a single monomer given by the WLC model [inverting main text Eq.~(2), as described in Ref.\cite{severino2019efficient}] and $\phi_{\textrm{F}}(f)$ can be written as [Eq.~(6), main text]:

\begin{equation}
\phi_{\textrm{F}}(f)=\frac{1}{2}\left(1+\frac{\sinh\left(\beta A(f)\right)}{\sqrt{e^{-4\beta B(f)}+\sinh^2\left(\beta A(f)\right)}}\right),
\label{eq:fraction_ssDNA2}
\end{equation}
with $\beta=1/k_{\rm B}\,T$ and A and B are given by:

\begin{eqnarray}
   A(f) & = & -\frac{\epsilon}{2}+\frac{1}{2} \int_0^{f}x_{\textrm{F}}(f')df'\,\, ;  \\ 
   B(f) & = & \gamma.
\end{eqnarray}

Note that $\phi_{\textrm{F}}$ corresponds to the fraction of unpaired bases $\phi$ defined in Ref.~\cite{bosco2014elastic}, and described by the following phenomenological formula:

\begin{equation}
\phi_{\textrm{F}}(f)=\frac{1}{1+e^{-(f-f^{\dagger})/\delta}}.
\label{eq:fraction_bosco}
\end{equation}

\begin{figure}[h!]
\centering
\includegraphics[width=\linewidth]{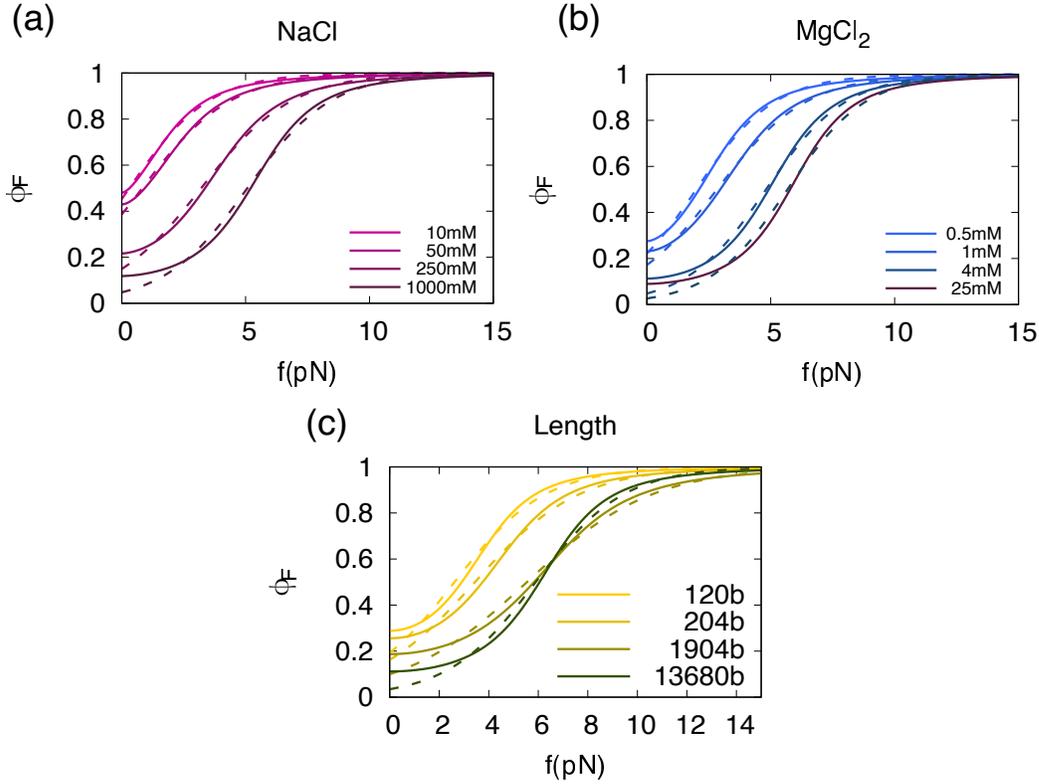}
\caption{{\bf Fraction of free bases as a function of the force.} Solid lines are obtained from the helix-coil model [Eq.~(\ref{eq:fraction_ssDNA2})], using the parameters of the fits (main text) while the dashed lines are fits of Eq.~(\ref{eq:fraction_bosco}) to the continuous lines for (a) varying NaCl, (b) \ce{MgCl2} concentrations and (c) molecular length.}
\label{fig:S:bosco1}
\end{figure}

\begin{figure}[h!]
\centering
\includegraphics[width=\linewidth]{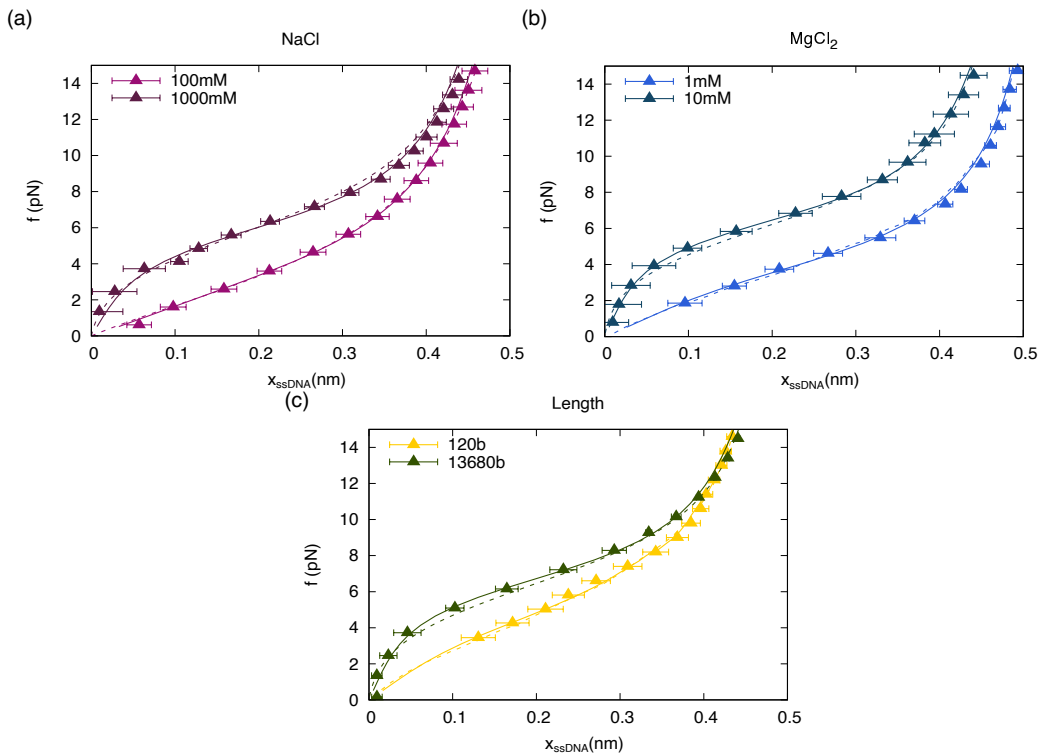}
\caption{{\bf FECS predicted from $\phi_{\textrm{F}}$ from Fig.~\ref{fig:S:bosco1}.} Theoretical FECs are obtained from Eq.~(\ref{eq:extension_model2}), with $\phi_{\textrm{F}}$ substituted from the helix-coil model [Eq.~(\ref{eq:fraction_ssDNA2})] (solid lines) and from the phenomenological formula [Eq.~(\ref{eq:fraction_bosco})] (dashed lines). Triangles show the experimental FECs for two conditions of varying (a) NaCl (b) \ce{MgCl2} and (c) length. Color code as in Fig.~\ref{fig:S:bosco1}. }
\label{fig:S:bosco3}
\end{figure}

In order to compare our model with this phenomenological description, we fitted Eq.~(\ref{eq:fraction_bosco}) ($f^{\dagger}$ and $\delta$) to $\phi_{\textrm{F}}$ curves [Eq.~(\ref{eq:fraction_ssDNA2})] with $\epsilon$ and $\gamma$ obtained from the fits of the helix-coil model to the experimental FECs [Fig.~3(a) and Fig.~4(b)-(c), main text]. In Fig.~\ref{fig:S:bosco1}, $\phi_{\textrm{F}}$ from the phenomenological fit (dashed line) is compared to the helix-coil model (continuous line), for varying NaCl conditions [panel (a)], \ce{MgCl2} conditions [panel (b)] and ssDNA length [panel (c)]. In each condition, both curves (continuous and dashed) follow a similar trend, and only differ significantly at low forces. Using Eq.~(\ref{eq:extension_model2}) we can also predict the FECs, as shown in Fig.~\ref{fig:S:bosco3}, where a good agreement between the phenomenological description (dashed lines) and the helix-coil model (continuous lines) is observed .  
Overall, the phenomenological formula reproduces successfully the experimental data, although some differences appear in the predictions for the fraction of free bases at low forces, especially at high salts.

The force drops faster from $\sim 2-3$ pN to zero in the predicted FECs from the helix-coil model than these obtained for the phenomenological formula, as observed in Fig.~\ref{fig:S:bosco3}. This is caused by the difference in the leading term in the extension [Eq.~(\ref{eq:extension_model2})]: at the limit $f\to0$, $x_\textrm{F}(f)\sim f$ (Hookean limit), which corresponds to a leading term in the extension of $\sim f^2$ for the phenomenological formula [Eq.~(\ref{eq:fraction_bosco})], and of $\sim f^3$ for the helix-coil model [Eq.~(\ref{eq:fraction_ssDNA2})].

The values for $f^{\dagger}$ and $\delta$ obtained from fitting the $\phi_{\textrm{F}}$ predicted from the helix-coil model are reported in Table~\ref{tab:table2}. These values are similar to those obtained from directly fitting the phenomenological formula Eq.~(\ref{eq:fraction_bosco}) to experimental FECs \cite{bosco2014elastic}, also shown in Table~\ref{tab:table2}. 
Note that the $\phi_{\textrm{F}}$ fitting procedure allows to estimate $f^{\dagger}$ and $\delta$ at low salt concentrations  (i.e. 10, 25 and 50mM NaCl) where the direct fitting to Eq.~(\ref{eq:fraction_bosco}) cannot be used (because secondary structure is formed at low forces and FECs do not present any clear force shoulder).

\begin{figure}[h!]
\centering
\includegraphics[width=\linewidth]{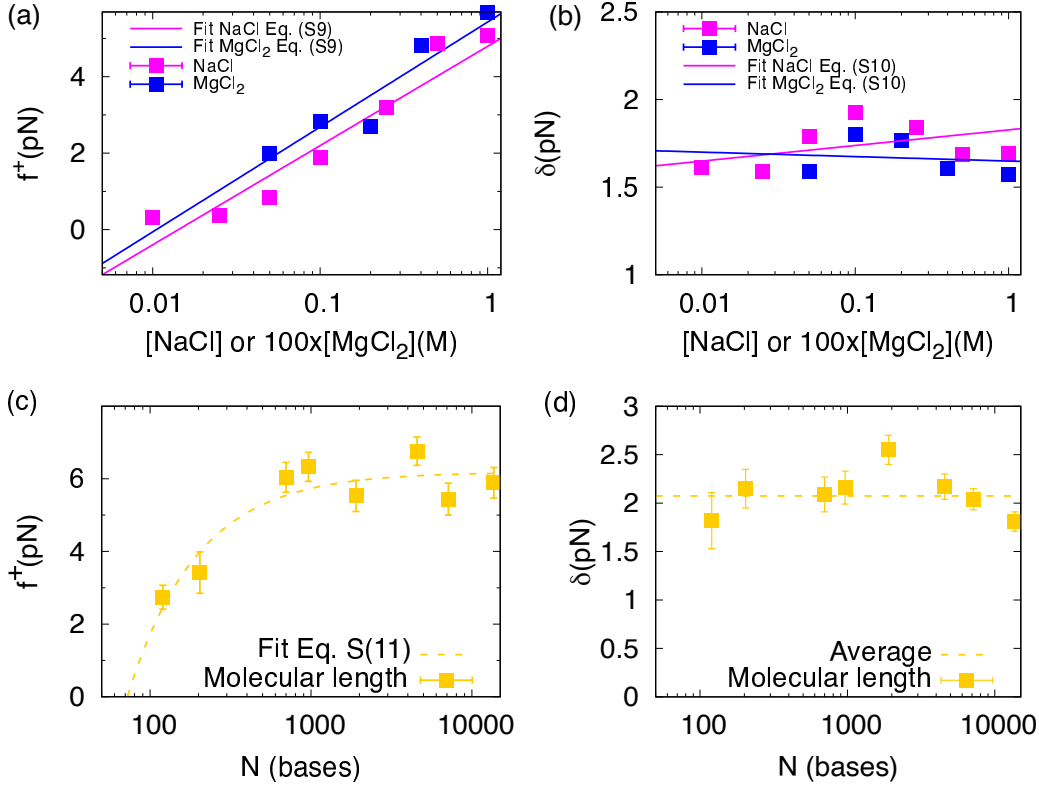}
\caption{{\bf Salt and length dependence of $f^\dagger$ and $\delta$.} The points show the fitting values of Eq.~(\ref{eq:fraction_bosco}) to $\phi_{\textrm{F}}$ [Eq.~(\ref{eq:fraction_ssDNA2})] for $f^\dagger$ and $\delta$ for varying (a)-(b) salt concentration and (c)-(d) molecular length. The lines in (a) and (b) are the fitting curves of Eqs.~(\ref{eq:fdag})-(\ref{eq:delta}), respectively, while in (c) and (d) they correspond to the fitting curves of Eq.~(\ref{eq:fdag_length}) and the average of the obtained values, respectively. }
\label{fig:S:bosco2}
\end{figure}

As shown in Fig.~\ref{fig:S:bosco2}(a) $f^{\dagger}$ is found to depend logarithmically with salt concentration:

\begin{equation}
    f^{\dagger}(\textrm{c})=a\log\left(\textrm{c}\right)+b,
    \label{eq:fdag}
\end{equation}
with $\textrm{c}=\textrm{[NaCl]}$ or $\textrm{c}=\textrm{[MgCl}_2$] (in M units). Fits to Eq.~(\ref{eq:fdag}) are shown in Fig.~\ref{fig:S:bosco2}(a) as dashed lines. The results from the fits are $a=1.7(2)$ pN and $b=4.9(5)$ pN in sodium and  $a=1.8(4)$ pN and $b=14(2)$ pN in magnesium. $\delta$ is also found to follow a logarithmic relation with salt concentration:

\begin{equation}
    \delta(\textrm{c})=\alpha\log\left(\textrm{c}\right)+\beta,
    \label{eq:delta}
\end{equation}
with $\textrm{c}=\textrm{[NaCl]}$ or $\textrm{c}=\textrm{[MgCl}_2$] (in M units). Fits to Eq.~\ref{eq:delta} are shown in Fig.~\ref{fig:S:bosco2}(b-c) as dashed lines. The results from the fits are $\alpha=-0.163(13)$ pN and $\beta=1.78(4)$ pN for sodium and $\alpha=-0.10(5)$ pN and $\beta=1.2(3)$ pN for magnesium.

Finally, the best fitting values for $f^{\dagger}$ and $\delta$ as a function of the molecular length are shown in Fig.~\ref{fig:S:bosco2}(c)-(d). While $\delta$ is constant [$\delta=2.04(8)$ pN], $f^{\dagger}$ follows a $1/N$ correction, with $N$ being the number of bases: 

\begin{equation}
    f^{\dagger}(N)=f^{\dagger}_0+\frac{f_{\textrm{N}}}{N}.
    \label{eq:fdag_length}
\end{equation}

The results from the fits are  are $f^{\dagger}_0=6.2(3)$ pN and $f_{\textrm{N}}=450(60)$ pN [Fig.~\ref{fig:S:bosco2}(c)]. Table~\ref{tab:table3} recompiles the formulae and values used for obtaining $f^{\dagger}$ and $\delta$ for varying length and salt concentration.

\begin{table}
\caption{\label{tab:table3}
Fitting parameters of Eq.~(\ref{eq:fraction_bosco}) obtained for each studied condition . }

\begin{tabular}{cccc}
\hline
\hline
Condition & Relation & & Values (pN) \\
\hline
Length  & $f^{\dagger}(N)=f^{\dagger}_0+\frac{f_{\textrm{N}}}{N}$ & & $f^{\dagger}_0=6.2(3)$, $f_{\textrm{N}}=450(60)$\\
(10mM \ce{MgCl2}) & $\delta=$constant & & $\delta=2.04(8)$  \\
\hline
 & & c=[NaCl] & $a=1.7(2)$, $b=4.9(5)$  \\
&  $f^{\dagger}(\textrm{c})=a\log\left(\textrm{c}\right)+b$ & & $\alpha=-0.163(13)$, $\beta=1.78(4)$ \\
Salt & & & \\
 & $\delta(\textrm{c})=\alpha\log\left(\textrm{c}\right)+\beta$ & c=[\ce{MgCl2}] & $a=1.8(4)$, $b=14(2)$  \\
& & & $\alpha=-0.10(5)$, $\beta=1.2(3)$ \\
\hline
\hline
\end{tabular}

\end{table}

\begin{table}[b]
\caption{\label{tab:table2}
Parameters ($\epsilon$ and $\gamma$) obtained from the fits of Fig.~3(a) and Fig.~4(a)-(b) (main text). The values of $f^{\dagger}$ and $\delta$ have been obtained by fitting Eq.~(\ref{eq:fraction_bosco}) to the obtained $\phi_F$ [Eq.~(\ref{eq:fraction_ssDNA2})] using the values of $\epsilon$ and $\gamma$ for each condition. The values of $f^{\dagger}$ and $\delta$ given in Ref.~\cite{bosco2014elastic} are also shown.}
{\footnotesize
\begin{tabular}{ccccccc}
 & $\epsilon$ (kcal/mol) & $\gamma$ (kcal/mol) & $f^{\dagger}$ (pN) & $\delta$ (pN) & $f^{\dagger}$ (pN)\footnotemark[2] & $\delta$ (pN)\footnotemark[2]  \\
\hline
\hline
H$_{120}$ & 0.035(17) & 0.69(3) & 2.61(2) & 1.83(2) & - & -  \\
H$_{204}$ & 0.066(17) & 0.62(2) & 3.44(2) & 2.10(2) & - & -  \\
H$_{700}$ & 0.174(18) & 0.57(2) & 6.05(2) & 2.05(2) & - & -  \\
H$_{964}$ & 0.189(19) & 0.59(3) & 6.33(2) & 2.13(2) & - & -  \\
H$_{1904}$ & 0.153(17) & 0.515(19) & 5.54(2) & 2.52(2) & - & -  \\
H$_{4454}$ & 0.208(18) & 0.541(19) & 6.77(2) & 2.15(2) & - & -  \\
H$_{7138}$ & 0.147(17) & 0.599(16) & 5.45(2) & 2.01(2) & - & -  \\
H$_{13680}$ & 0.166(17) & 0.643(16) & 5.90(2) & 1.77(2) & 5.68 & 1.43   \\
\hline
[NaCl](mM) & & & & & & \\
10\footnotemark[1]  & -0.12 & 0.83 & -2.05(3) & 2.46(2) & - & -   \\
25\footnotemark[1]  & -0.07 & 0.79 & -0.95(2) & 2.40(2) & - & -   \\
50\footnotemark[1]  & -0.03 & 0.77 & -0.03(2) & 2.18(2) & - & -   \\
100 & -0.018(17) & 0.77(2) & -0.31(2) & 2.18(2) & 0.44 & 2.09 \\
250 & 0.039(16) & 0.69(2) & 2.28(2) & 2.06(2) & 3.6 & 1.75 \\
500 & 0.115(17) & 0.67(3) & 4.59(2) & 1.83(2) & 4.85 & 1.64  \\
1000 & 0.125(17) & 0.69(2) & 4.93(2) & 1.77(2) & 5.32 & 1.52 \\
\hline
[MgCl$_2$](mM) & & & & & & \\
0.5 & -0.01(2) & 0.81(3) & 0.20(2) & 1.87(2) & 0.62 & 1.96 \\
1 & 0.03(2) & 0.73(3) & 1.78(2) & 1.95(2) & 2.27 & 2.07  \\
2 & 0.02(2) & 0.74(3)& 1.45(2) & 2.03(2) & 3.31 & 1.58 \\
4 & 0.11(2) & 0.69(2) & 4.57(2) & 1.71(2) & 4.90 & 1.61 \\
10 & 0.15(2) & 0.675(17) & 5.61(2) & 1.65(2) & 5.68 & 1.43  \\
\hline
\hline
\end{tabular}}
\end{table}
\footnotetext[1]{$\epsilon$ is obtained from Eq.~(14) [magenta dashed line in Fig.~4(c)] and $\gamma$ from Eq.~(15) [magenta dashed line in Fig.~4(d)] (main text).}
\footnotetext[2]{Values reported in Ref.~\cite{bosco2014elastic}.}

\clearpage
\newpage

\bibliographystyle{unsrt}
\bibliography{bibliography.bib}

\end{document}